\documentclass[apj, onecolumn]{emulateapj}

\usepackage{graphicx}
\usepackage{epsfig}
\usepackage{natbib}
\usepackage{amssymb,amsmath}
\newcommand{\vel}{ {\bf v} }

\newcommand{\bB}{{\bf B}}
\newcommand{\bk}{{\bf k}}

\newcommand{\bfaltg}{\boldsymbol{\mathcal G}}
\newcommand{ \er }{ {\rm\bf e}_r}

\newcommand{\misfit}{ \mathcal{I} }

\newcommand{ \altg }{ {\mathcal G}}
\newcommand{ \noise }{ {\mathcal N}}

\newcommand{ \curl }{ {\mbox{\boldmath$\times$}} }
\newcommand{ \fwd }{ {\mbox{\boldmath$\Phi$}} }
\newcommand{ \adj }{ {\mbox{\boldmath$\Phi^\dagger$}} }
\newcommand{ \baradj }{ {\mbox{\boldmath$\bar\Phi^\dagger$}} }
\newcommand{ \barfwd }{ {\mbox{\boldmath$\bar\Phi$}} }
\newcommand{ \lamv }{ {\mbox{\boldmath$\lambda$}} }

\newcommand{ \bnabla }{ {\mbox{\boldmath$\nabla$}} }
\newcommand{ \bxi }{ {\mbox{\boldmath$\xi$}} }
\newcommand{ \bzeta }{ {\mbox{\boldmath$\eta$}} }

\newcommand{ \bx }{ {\bf x} }
\newcommand{ \cc }{ {\mathcal C} }

\begin{document}
\title{THE ADJOINT METHOD APPLIED TO TIME-DISTANCE HELIOSEISMOLOGY}

\author{Shravan M. Hanasoge\altaffilmark{1,2}, Aaron Birch\altaffilmark{3}, Laurent Gizon\altaffilmark{1,4}, \& Jeroen Tromp\altaffilmark{2,5}}
\altaffiltext{1}{Max-Planck-Institut f\"{u}r Sonnensystemforschung, Max Planck Stra{$\beta$}e 2, 37191 Kaltenburg-Lindau, Germany}
\altaffiltext{2}{Department of Geosciences, Princeton University, Princeton, NJ 08544, USA}
\altaffiltext{3}{NorthWest Research Associates, Colorado Research Associates, Boulder, CO 80301, USA}
\altaffiltext{4}{Georg-August-Universit\"{a}t, Institut f\"{u}r Astrophysik, Friedrich-Hund-Platz 1, D-37077 G\"{o}ttingen, Germany}
\altaffiltext{5}{Program in Applied \& Computational Mathematics, Princeton University, Princeton, NJ 08544, USA}

\begin{abstract}
For a given {\it misfit function}, a specified optimality measure of a model,
its gradient
describes the manner in which one may alter properties of the system to march towards a stationary point. 
The adjoint method, arising from partial-differential-equation-constrained optimization, describes a means of extracting
derivatives of a misfit function with respect to model parameters through finite computation. 
It relies on the accurate calculation of wavefields that are driven by two types of sources, namely
the average wave-excitation spectrum, resulting in the {\it forward wavefield}, and differences between predictions and observations, 
resulting in an {\it adjoint wavefield}. All sensitivity kernels relevant to a given measurement emerge directly 
from the evaluation of an interaction integral involving these wavefields. The technique facilitates computation of sensitivity kernels (Fr\'{e}chet derivatives)
relative to three-dimensional heterogeneous background models, thereby paving the way for non-linear iterative inversions.
An algorithm to perform such inversions using as many observations as desired is discussed. 
\end{abstract}

\section{Introduction}
Diffuse seismic wavefields, present in a variety of media, such as stars and planets, may be created by the action of stochastic sources of wave excitation, where
source location, amplitude and phase are random variables. Without knowledge of the exact realization of all relevant sources of wave 
excitation, raw time series of wavefield velocities contain no useful seismic information. However, it was discovered that seismically relevant data were contained 
in time-averages over many source realizations of second-order correlations of wavefield velocities in the Sun (\citet{duvall}; for noise tomography, see, e.g., \citet{shapiro04}).
These correlations contain components of noise, whose standard deviation 
diminishes as the inverse square root of the temporal length of averaging, which is the consequence of
wave excitation by a stationary random process \citep[see, e.g.,][]{gizon02,gizon04}. In the Sun, turbulent convection, driven by radiative thermal losses at the
surface, is the cause of wave generation. This random process is adequately
represented by a stationary and laterally homogeneous random process
\citep{gizon04}. Typically, the correlation time of solar convection
(granulation) is 10~min and the correlation length is 1~Mm. While the
correlation time is on the order of the wave period, the correlation length
is smaller than the wavelength and thus the assumption of spatially
uncorrelated sources is reasonable. The response of the Sun to excitation
by turbulent convection produces a power spectrum that peaks near 3 mHz.

\citet{woodard} and \citet{gizon02} were among the first to utilize these ideas towards the construction of a theoretical description of helioseismic measurements. A 
prescription to compute sensitivity kernels and model excitation noise was described by \citet{gizon02, gizon04}. 
Various authors, e.g., \citet{birch}, \citet{birch_gizon_07} and \citet{jason07}, subsequently used this
theory to derive sensitivity kernels for flows and sound-speed perturbations for translationally-invariant 
(laterally homogeneous) background models. The basic recipe described in \citet{gizon02} to compute travel-time
sensitivity kernels for randomly excited waves was general; however no
attempt was made to relate this method to the adjoint method, which enables
computation of kernels for heterogeneous background models using numerical wave
simulations.

The adjoint method has a long history \citep{lions71} and is widely used in fluid control \citep[e.g.,][and references therein]{bewley01, giles.adjoint.00}, airfoil optimization \citep[e.g.,][]{jameson88}, meteorology \citep[e.g.,][]{dimet86,talagrand87} and terrestrial seismology \citep[e.g.,][]{tarantola84, tromp05,tromp10}. Real-world
optimization problems are typically functions of large numbers of parameters,  ill posed and computationally expensive. For instance, one may envisage the difficulty in minimizing
drag due to flow over an airfoil or seeking a model of Earth's interior that optimally fits observed seismograms, simply due to large number of ways one may alter the system. What parameters should one 
vary in order to achieve optimality? It is evident that the gradient of the misfit function
with respect to various parameters tells us how to march towards a stationary point, i.e., a point at which the derivative of a quantity vanishes. The adjoint method
provides an algorithm to compute Fr\'{e}chet derivatives and hence the gradient with relatively small computational expense.

In this article we extend the adjoint method to the computation of helioseismic sensitivity kernels relative to arbitrarily heterogeneous background models. The complexity of equations in the presence of strong
lateral inhomogeneity is such that evaluation of kernels must proceed by computational means. A remarkable outcome of allowing for lateral (horizontal) variations in the background model
 is the ability
to compute vector kernels for magnetic fields. Since the perturbation induced by fields scales as $O(|{\bf B}|^2)$, where ${\bf B}$ is the background field, and the action of the Lorentz force is
anisotropic, it is non-trivial to derive magnetic field kernels about a 1D solar model. However, if we were to linearize around a 3D background model that contains an embedded field ${\bf B}$,
kernels describing shifts in helioseismic measurements due to small vector variations in the magnetic field emerge naturally. 

Evaluating kernels in the context of helioseismology requires the computation of six wavefields per measurement. 
Losing the luxury of being able to translate kernels from one horizontal position to another 
therefore comes at a stiff computational price, since one must, in principle, evaluate kernels around each observational pixel, an impossible
feat in helioseismology owing to the vast numbers of observations. Michelson Doppler Imager \citep[MDI;][]{scherrer95} records velocities at approximately 1 million points on the solar photosphere, and
with the advent of Solar Dynamics Observatory (SDO), Helioseismic and Magnetic Imager (HMI) now captures velocities at more than 16 million pixels every 45 seconds. Computing kernels at all these points is 
neither computationally feasible nor is it clear that there is sufficient independent information to require such a massive calculation. Consequently, we introduce the concept of ``master pixels", a 
finite constellation of points which we consider interesting enough to invest this sizeable computational effort. However, once a number of these pixels have been chosen, 
{\it every} cross correlation measurement, one of whose antennae is a master pixel, may be utilized in the inversion without affecting computational cost \citep{tromp10}.

In this article, we shall primarily discuss the mathematical underpinnings of the adjoint method and its applicability to helioseismology. A computational algorithm to implement
the analysis is described. Perturbations, such as sunspots, are significant deviations from
the quiet Sun and shifts in helioseismic measurements in and around sunspots are substantial and unlikely to scale linearly with perturbation strength (when measured relative to the quiet Sun).
A means of carrying out iterative 
inversions in such situations is described. With the increasing availability of computational resources, 
demand for greater accuracy in the interpretation of helioseismic measurements and the advent of higher-quality observations, the introduction of such a technique is thought to be timely.

\section{Governing equations of the helioseismic wavefield}\label{equations}
We start with a background state in magneto-static equilibrium, described by:
\begin{eqnarray}
\bnabla p &=& \rho {\bf g}  + (\bnabla\curl{\bf B})\curl{\bf B},\label{equili}\\
\bnabla\cdot{\bf g} &=& -4\pi G\rho,
\end{eqnarray}
where $p$ is the background pressure, $\rho$ the density, $\bB$ the magnetic field, ${\bf g} = - g \er$ gravity, $G$ the universal gravitational constant, and $\er$ the radially outward unit vector \citep[e.g.,][]{ostriker67,goedbloed2004,cameron07}. In this formalism, we consider background flows ($\vel$) to be too weak to contribute significantly towards maintaining 
equilibrium, and hence we neglect advection-related forces in equation~(\ref{equili}). This implies that $ ||\vel|| \ll \sqrt{g L}$, where $L$ is the characteristic flow length scale, and that Lorentz forces
are primarily balanced by pressure gradients and gravity.
However any flows that are present must satisfy the continuity equation, and we require 
therefore that $\bnabla\cdot(\rho\vel) = 0$.
The magnetic permeability constant $4\pi\mu_0$ has been absorbed into the definition of the field. We do not keep rotation terms in the force balance equation because Coriolis and centrifugal forces are five orders in
magnitude smaller than surface gravity. We also invoke the Cowling approximation, allowing us to ignore changes in the gravitational potential induced by wave motions.
Small-amplitude wave propagation in a magnetic environment is described by the following dynamical wave operator (in temporal Fourier space; see Appendix~\ref{conventions} for the convention)
\begin{eqnarray}
\boldsymbol{\mathcal L}\bxi&=&-\omega^2\rho\bxi-2i\omega\rho\vel\cdot\bnabla\bxi -i\omega\rho\Gamma\bxi - \bnabla(c^2\rho\bnabla\cdot\bxi)  -  \bnabla(\bxi\cdot\bnabla p) +{\bf g} \bnabla\cdot(\rho\bxi)\nonumber\\
&-&(\bnabla\curl\bB) \curl [\bnabla\curl(\bxi \curl \bB)] - \{\bnabla\curl[\bnabla\curl(\bxi \curl \bB)]\} \curl \bB,\label{waveop.helio}
\end{eqnarray}
where $\bxi$ is the displacement vector and $c$ is the background sound speed. In order, terms on the right side denote acceleration (first term),
flow advection, wave damping, pressure restoring forces (the term with $c^2\rho$), buoyancy (the next two terms) and magnetic Lorentz force (the final two) respectively.
We
assume that the upper boundary is placed far away from the solar photosphere and the wavefield satisfies zero-Dirichlet conditions (all fluctuations are zero on this bounding surface). The 
entire solar interior is enclosed within this volume.
Following \citet{gizon02} and \citet{birch}, we mimic the complex frequency dependence of wave damping in the Sun by including the term
$-i\omega \Gamma\bxi$, where $\Gamma$ is the damping rate.
We neglect second-order flow terms such as $\vel\cdot\bnabla(\vel\cdot\bnabla\bxi)$; this 
is a reasonable approximation when the velocities roughly satisfy $||\vel|| \ll \{\omega L,  c, \sqrt{g L}\} $, where $\omega$ is the characteristic wave frequency. It may be verified that this inequality is satisfied for most solar phenomena \citep[e.g.,][]{birch_gizon_07}. The full wave equation is given by $\boldsymbol{\mathcal L}\bxi = {\bf S}$, where ${\bf S}(\bx,\omega)$ is a source term.

Flows and damping do not follow directly from the equilibrium equations. The emergence of wave damping 
is not well understood, and is thought to be due to a combination of the action of turbulence and radiation \citep[e.g.,][]{duvall98, korzennik}; we are unable
to realistically account for these phenomena and are therefore forced to introduce phenomenological damping terms.
Solar flows, as discussed previously, are typically weak perturbations to the background. Further, constructing a background model with flows 
and magnetic fields is a remarkably difficult task \citep[e.g.,][]{belien}. Such practical considerations have led us to introduce these
terms in an ad-hoc fashion.

\section{Minimizing misfit}
A common optimization problem in helioseismology is that of reducing differences between observed and predicted travel times. 
Cross-correlation amplitudes, which depend quasi-linearly on properties of the background model, are commonly measured but
not typically used in inversions; conceptually, one may include these in the misfit with no additional effort \citep[e.g.,][]{gee92,fichtner08,bozdag11}.  

A convenient choice for the misfit function is the $L_2$ norm of these differences summed over a number of observation points 
\begin{equation}
\misfit' = \frac{1}{2}\sum_{q,q'} \noise_{qq'}[\tau^{(n)}_{q} - \tau^{\mathrm o}_{q}][\tau^{(n)}_{q'} - \tau^\mathrm{o}_{q'}],\label{misfit.tt.only}
\end{equation}
where $\tau^{\mathrm o}_{q}$ is the observed travel time, $\tau^{(n)}_q$ the predicted analog with (current) background model $n$, 
specified at points $q,q'$, and $\noise_{qq'}$ the inverse of the noise covariance
between the two sets of measurements, assumed to be chi-squared distributed \citep{gizon04}. Here the noise-covariance model is assumed to be stationary under
changes of the background model, i.e., ${\noise_{qq'}}$ does not change with iteration. Partial-differential-equation constrained optimization is the technique 
of minimizing this misfit with respect to a governing wave equation,
\begin{equation}
\misfit = \frac{1}{2}\sum_{q,q'} \noise_{qq'}[\tau^{(n)}_{q} - \tau^o_{q}][\tau^{(n)}_{q'} - \tau^o_{q'}] - \int_\odot d\bx\int  d\omega~\lamv\cdot(\boldsymbol{\mathcal L}\bxi -{\bf S}),\label{misfit.orig}
\end{equation}
where $\lamv$ is a Lagrange multiplier and the integration proceeds over all space $\bx$ and frequency $\omega$.
As, e.g., \citet{woodard} and \citet{gizon02} realized, a first step towards formal interpretation of measurements is to create functionals linking cross correlations and 
travel times to the input displacement field, i.e., to establish a relation of the form $\tau^{(n)}_{q} = \tau^{(n)}_{q}(\bxi)$. 
For now, we choose to represent this in an abstract fashion, and in subsequent sections move towards greater detail. Let us posit that
a change in misfit~(\ref{misfit.tt.only}) may be written as
\begin{equation}
\delta\misfit' = \int_\odot d\bx\int d\omega~{\bf f}^\dagger \cdot \delta\bxi,\label{gen.var}
\end{equation}
where ${\bf f}^\dagger$ is a function that connects variations in displacement field $\delta\bxi$, to those of travel-time misfit $\delta\misfit'$.
Now changes in the misfit associated with the constrained problem~(\ref{misfit.orig}) may be written as
\begin{equation}
\delta\misfit = \int_\odot d\bx~\int d\omega~{\bf f}^\dagger \cdot \delta\bxi - \int_\odot d\bx\int d\omega~[\delta\lamv\cdot(\boldsymbol{\mathcal L}\bxi -{\bf S}) + \lamv\cdot\delta\boldsymbol{\mathcal L}\,\bxi + \lamv\cdot\boldsymbol{\mathcal L}\,\delta\bxi],\label{del.misf}
\end{equation}
upon invoking~(\ref{gen.var}) and setting $\delta{\bf S} = {\bf 0}$. If the forward displacement field were to satisfy $\boldsymbol{\mathcal L}\bxi ={\bf S}$, and we were able to eliminate terms involving $\delta\bxi$, 
then changes in the misfit would be functions only of $\lamv$, $\bxi$ and the perturbed wave operator, which depends only on background properties such as sound speed, magnetic fields, density, etc.
Now in order to accomplish this, we need to first be able to free $\delta\bxi$ from the action of the operator in the third term of equation~(\ref{del.misf}).
The property of adjointness or duality is central to such a manipulation. An operator ${\mathcal O}$ is said to be {\it self-adjoint} if it satisfies 
\begin{equation}
\int_\odot d\bx~ \lamv\cdot{\mathcal O}\bxi = \int_\odot d\bx~ \bxi\cdot{\mathcal O}\lamv.
\end{equation}
For the boundary conditions chosen here, it may be demonstrated that the 
ideal MHD operator, which contains no flow or dissipation terms, is an example \citep[e.g.,][]{goedbloed2004}. However, the non-ideal operator~(\ref{waveop.helio}) is not self-adjoint and obeys
\begin{equation}
\int_\odot d\bx~ \lamv\cdot\boldsymbol{\mathcal L}\bxi = \int_\odot d\bx~ \bxi\cdot\boldsymbol{\mathcal L}^{\dagger}\lamv,
\end{equation} 
where $\boldsymbol{\mathcal L}^\dagger$, defined as adjoint to~(\ref{waveop.helio}), is given by (see appendix~\ref{sec.reciprocity})
\begin{eqnarray}
\boldsymbol{\mathcal L}^\dagger\bxi&=&-\omega^2\rho\bxi -i\omega\rho\Gamma\bxi+2i\omega\rho\vel\cdot\bnabla\bxi - \bnabla(c^2\rho\bnabla\cdot\bxi  +  \bxi\cdot\bnabla p) +{\bf g} \bnabla\cdot(\rho\bxi)
 \nonumber\\
&&-\left[(\bnabla\curl\bB) \curl \{\bnabla\curl(\bxi \curl \bB)\} + \{\bnabla\curl[\bnabla\curl(\bxi \curl \bB)]\} \curl \bB\right].\label{waveop.helio.adjoint}
\end{eqnarray}
 The only difference between operators~(\ref{waveop.helio}) and~(\ref{waveop.helio.adjoint}) is that of a reversal in sign of the background flow term ($\vel$ flips sign).
Thus the following term may be rearranged such that
\begin{equation}
\int_\odot d\bx \int d\omega~\lamv\cdot\boldsymbol{\mathcal L}\,\delta\bxi = \int_\odot d\bx\int  d\omega~\delta\bxi\cdot\boldsymbol{\mathcal L}^\dagger\,\lamv,
\end{equation}
where $\boldsymbol{\mathcal L}^\dagger$, the adjoint (or dual) operator, acts on Lagrange multiplier $\lamv$ and $\delta\bxi$ has been effectively freed.
Now, we choose $\lamv$ so as to satisfy the differential equation
\begin{equation}
\boldsymbol{\mathcal L}^\dagger\lamv - {\bf f}^\dagger = {\bf 0},
\end{equation}
leaving an elegant and simple connection between the variation in misfit and model parameters:
\begin{equation}
\delta\misfit = -\int_\odot d\bx\int d\omega~\lamv\cdot\delta\boldsymbol{\mathcal L}\bxi.
\end{equation}
Since $\boldsymbol{\mathcal L}$ depends solely on background properties (denoted collectively as $\{\beta_s\}$), variations in
the operator may be represented as effective functions of $\delta\beta_s$
\begin{equation}
\lamv\cdot\delta\boldsymbol{\mathcal L}\,\bxi = \left(\lambda_i\,\,\frac{\partial{\mathcal L}_{ij}}{\partial\beta_{s}}\,\,\xi_j\right)\,\delta\beta_s,
\end{equation}
where Einstein's summation convention is employed and $\boldsymbol{\mathcal L} = \{{\mathcal L}_{ij}\}$ is a second-order tensor. Properties $\beta_s$
in this theory are regarded as being functions only of space; this allows us to define a kernel as
\begin{equation}
K_s = -\int d\omega~\left(\lambda_i\,\frac{\partial{\mathcal L}_{ij}}{\partial\beta_{s}}\,\xi_j\right),
\end{equation}
leading to
\begin{equation}
\delta\misfit = \int_\odot d\bx ~\sum_s K_s\, \delta\beta_s,\label{delta1}
\end{equation}
which tells us how to simultaneously solve the inverse problem for all relevant helioseismic quantities:
\begin{equation}
\delta\misfit = \int_\odot d\bx~\left(K_{\rho}~ \delta\rho + K_{c^2}~ \delta c^2  + {\bf K}_{\vel}\cdot \vel + {\bf K}_{\bB}\cdot\delta \bB\right).\label{delta2}
\end{equation}
Note we have not written out a kernel for pressure since it may be determined by considering variations in the equilibrium equation~(\ref{equili}). For a more
detailed treatment, please see section~\ref{sens.kernels}, equation~(\ref{perturb.equili}).
When performing an iterative inversion, it is evident from equation~(\ref{delta1}) that by choosing $\delta\beta_s = -\epsilon_s K_s$, where $\epsilon_s > 0$ is
a small constant, we arrive at,
\begin{equation}
\delta\misfit = -\int_\odot d\bx ~\sum_s \epsilon_s K^2_s < 0.
\end{equation}
This is the principle of the steepest descent method. 
More sophisticated inverse algorithms such as the conjugate-gradient method, which uses previous and current 
gradients to construct the model update at a given iteration level, may be more relevant. Pre-conditioning, a technique applied to improve the condition number, may also be implemented.
The determination of $\epsilon$ is also non-trivial, requiring a ``line search'' to determine an optimal value (whereas a crude way is to simply set it to some small
value, such as 0.02). Thus, we alter the background state by amounts directly proportional to the Fr\'{e}chet derivative, i.e., we perform the following updates:
\begin{eqnarray}
c^2 &\rightarrow& c^2 - \epsilon_c~ K_{c^2}, \nonumber\\
\rho &\rightarrow& \rho - \epsilon_\rho~ K_{\rho}, \nonumber\\
\vel &\rightarrow& \vel - \epsilon_\vel~ {\bf K}_{\vel},\nonumber \\
\bB &\rightarrow& \bB - \epsilon_\bB~ {\bf K}_{\bB}.\label{update.eqs}
\end{eqnarray}
The preceding set of equations describes in generality how to pose the helioseismic inverse problem; translational invariance
is a specific case of this formalism. 

\section{Measurement functionals} 
Thus far, we have very generally described the underpinnings of the adjoint method; from this point on, we focus on the primary measurement in time-distance
helioseismology: cross correlations.
Since we are interested in determining the gradient of the misfit function based on travel times (Eq.~[\ref{misfit.tt.only}]), we must 
both appreciate how travel times are computed and quantify their variation with respect to changes in model parameters. Varying equation~(\ref{misfit.tt.only}), we have
\begin{eqnarray}
\delta\misfit' &=& \frac{1}{2}\sum_{q,q'} \noise_{qq'}\,[\Delta\tau^{(n)}_{q'}\delta\tau_{q} + \Delta\tau^{(n)}_{q}\,\delta\tau_{q'}],\\
\delta\misfit' &=& \sum_{q,q'} \frac{1}{2}(\noise_{qq'} + \noise_{q'q})\,\Delta\tau^{(n)}_{q'}\,\delta\tau_{q},\\
&=& \sum_{q} b^{(n)}_q\delta\tau_{q},\label{temp.misfit}\\
b^{(n)}_q &=& \sum_{q'} \frac{1}{2}(\noise_{qq'} + \noise_{q'q})\,\Delta\tau^{(n)}_{q'},
\end{eqnarray}
where $\Delta\tau^{(n)}_{q} = [\tau^{(n)}_{q} - \tau^o_{q}]$. We do not place the iteration superscript $n$ over the variation
in travel time $\delta\tau_q$ because this term implicitly depends on the background, which evolves with each iteration.  
We apply the following definition of travel time \citep[appendix A of][]{gizon02}
\begin{equation}
\delta\tau = \int_0^T dt'~W_{\alpha\beta}(t')~\delta\cc_{\alpha\beta}(t'),\label{var_tt}
\end{equation}
where $W_{\alpha\beta}$ is a weight function and $\delta\cc_{\alpha\beta}(t')$ the deviation in the cross correlation,
$\alpha,\beta$ are measurement pixel locations, and $T$ is the length of the temporal window.
Following \citet{woodard} and \citet{gizon02}, we begin by defining the cross correlation
\begin{equation}
\cc_{\alpha\beta}(t) = \frac{1}{T} \int_0^T \phi(\bx_\alpha,t')~\phi(\bx_\beta,t+t') ~dt',\label{cc.eq}
\end{equation} 
where $\phi(\bx, t)$ is the line-of-sight projected wave velocity measured at spatial point $\bx$ at the solar photosphere.  
Appropriate filters and point-spread-function contributions are assumed to have already been incorporated into the definition of $\phi(\bx, t)$. Transformed
into temporal Fourier space, this becomes
\begin{equation}
\cc_{\alpha\beta} = \frac{1}{T}\,\phi^*(\bx_\alpha,\omega)\phi(\bx_\beta,\omega).\label{cc.omega}
\end{equation} 

Let Green's tensor for the system of differential equations, denoted by ${\bf G}(\bx,\bx',\omega)$, satisfy
\begin{equation}
\boldsymbol{\mathcal L}{\bf G} = \delta(\bx-\bx') ~{\bf I},\label{greenseq}
\end{equation}
where $\bx$ is termed the ``receiver" and $\bx'$, ``the source", and ${\bf I} = \{\delta_{ij}\}$. Similarly, we define the adjoint Green's tensor via
\begin{equation}
\boldsymbol{\mathcal L}^\dagger{\bf G}^\dagger = \delta(\bx-\bx') ~{\bf I}.\label{greenseq.adj}
\end{equation}
Thus for an arbitrary source distribution ${\bf S}(\bx',\omega)$, the wavefield in
temporal Fourier domain is given by
\begin{equation}
\bxi(\bx,\omega) = \int_\odot d\bx'~{\bf G}(\bx,\bx',\omega)\cdot {\bf S}(\bx',\omega),
\end{equation}
and in time domain,
\begin{equation}
\bxi(\bx,t) = \int_\odot d\bx'\int dt'~{\bf G}(\bx,\bx',t-t')\cdot {\bf S}(\bx',t').
\end{equation}
Similar relations apply to the adjoint wavefield.
In order to reduce
notational burden, we discontinue explicitly writing the $\omega$ dependence, i.e., only source and receiver locations will be included when stating Green's function.
In analyses that follow, we shall repeatedly switch positions of the source and receiver. Green's functions in the case of a switched source-receiver pair
satisfies the following reciprocity relation (see appendix~\ref{sec.reciprocity})
\begin{equation}
{\bf G}^\dagger(\bx',\bx) = {\bf G}^T(\bx,\bx') \label{recip.main}.
\end{equation}

Observations are typically highly processed versions of the raw solar vector velocity field, subjected to point spreading and phase-speed filtering, line-of-sight projection, etc. 
Following \citet{gizon02}, we introduce vector $\altg_j$ to denote Green's function for the filtered, line-of-sight projected velocity
\begin{equation}
\altg_j(\bx,\bx') = {\mathcal F}(\bx,\omega) * l_i(\bx) G_{ij}(\bx,\bx'),\label{def.altg}
\end{equation}
where the convolution is spatio-temporal, ${\bf\hat l} = \{l_i(\bx)\}$ is the unit line-of-sight projection vector, and ${\mathcal F}(\bx,\omega)$ contains all filter terms and the transformation between
displacement and observed wavefield velocity. Applying equation~(\ref{recip.main}) to~(\ref{def.altg}), we may define the reciprocal filtered Green's function
\begin{equation}
\altg^\dagger_j(\bx',\bx) = {\mathcal F}(\bx,\omega) * l_i(\bx) G^\dagger_{ji}(\bx',\bx).\label{def.altg.2}
\end{equation}
Note that $\altg_j(\bx,\bx') \equiv \altg^\dagger_j(\bx',\bx)$, since all we do is to replace $G_{ij}(\bx,\bx')$ by its adjoint counterpart $G^\dagger_{ji}(\bx',\bx)$, to which
it is identically equal.

The cross correlation written in terms of Green's tensors, driven by the source $S_k(\bx,\omega)$, where $k$ is the direction of the dipolar source, is
\begin{equation}
\cc_{\alpha\beta} = \frac{1}{T}\int_\odot d\bx'\int_\odot d\bx{''} ~\altg^*_i(\bx_\alpha,\bx') ~\altg_j(\bx_\beta,\bx{''})~S^*_i(\bx',\omega)~S_j(\bx{''},\omega).
\end{equation}
Measured cross correlations are typically averaged over a large number of source-correlation times, allowing us to treat it as an ensemble average over many source
realizations. In other words, we consider a limit cross correlation that has detached itself from detailed properties of source action and is sensitive only to the statistical quantity 
$\langle S^*_i(\bx',\omega)~S_j(\bx{''},\omega)\rangle$, where the angled brackets denote ensemble averaging \citep[e.g.,][]{woodard,gizon02,tromp10}. In order to render this theory computable, we explicitly assume that sources at disparate spatial points are spatially uncorrelated,
allowing us to write
\begin{equation}
\langle S^*_i(\bx',\omega)~S_j(\bx{''},\omega)\rangle = \delta(\bx'-\bx'')~{\mathcal P}_{ij}(\bx',\omega),
\end{equation}
where ${\mathcal P}_{ij}$ encapsulates the average temporal power spectrum, correlations between different dipole sources and the spatial distribution of source amplitudes.
Thus the limit cross correlation becomes
\begin{equation}
\langle\cc_{\alpha\beta}\rangle = \frac{1}{T}\int_\odot d\bx'~\altg^*_i(\bx_\alpha,\bx') ~\altg_j(\bx_\beta,\bx{'})~{\mathcal P}_{ij}(\bx',\omega).\label{crosscorr.eq}
\end{equation}

Consider a variation in the cross correlation
\begin{eqnarray}
\langle\delta\cc_{\alpha\beta}\rangle &=& \frac{1}{T}\int_\odot d\bx'~[\altg^*_i(\bx_\alpha,\bx') ~\delta\altg_j(\bx_\beta,\bx{'}) + \delta\altg^*_i(\bx_\alpha,\bx') ~\altg_j(\bx_\beta,\bx{'})]~{\mathcal P}_{ij}\label{vareq},
\end{eqnarray} 
where we have chosen to neglect changes in properties of the power spectrum, i.e., $\delta{\mathcal P}_{ij}(\bx',\omega) =0$.
We invoke the first-Born approximation to describe variations in Green's tensor due to changes in properties of the background medium
\begin{equation}
\boldsymbol{\mathcal L}~\delta {\bf G} =  -\delta{\boldsymbol{\mathcal L}} ~{\bf G}.
\end{equation}
Using Green's identity, we recover the following expression for $\delta G_{ij}(\bx,\bx')$
\begin{equation}
\delta G_{ij}(\bx,\bx') = - \int_\odot d\bx{''}~ G_{ik}(\bx,\bx{''}) ~[\delta\boldsymbol{\mathcal L}~ {\bf G}(\bx{''},\bx')]_{kj},
\end{equation}
where the spatial coordinate in $\delta\boldsymbol{\mathcal L}$ is $\bx{''}$.
Finally, we have
\begin{equation}
\delta\altg_j(\bx,\bx') = {\mathcal F}(\bx,\omega) * [l_i~\delta G_{ij}] = -\int_\odot d\bx{''} ~\altg_k(\bx,\bx{''}) ~[\delta\boldsymbol{\mathcal L}~{\bf G}(\bx{''},\bx')]_{kj},
\end{equation}
where $\delta\boldsymbol{\mathcal L}$ is a function of $\bx{''}$ and the filter ${\mathcal F}(\bx, \omega)$ acts only on $l_i(\bx)\,G_{ik}(\bx,\bx{''})$.
Considering only the first term in the variation of the cross correlation in equation~(\ref{vareq}), we have
\begin{equation}
\langle\delta\cc^1_{\alpha\beta}\rangle =  -\frac{1}{T}\int_\odot d\bx\int_\odot d\bx'~[\altg^*_i(\bx_\alpha,\bx') ~\altg_k(\bx_\beta,\bx)] ~[\delta\boldsymbol{\mathcal L}~{\bf G}(\bx,\bx')]_{kj}~{\mathcal P}_{ij}.
\end{equation}
Rearranging the integration order,
\begin{eqnarray}
\langle\delta\cc^1_{\alpha\beta}\rangle &=&  -\frac{1}{T}\int_\odot d\bx~\altg_k(\bx_\beta,\bx) ~\left\{\delta{\mathcal L}_{kp}~\left[ \int_\odot d\bx'~{G}_{pj}(\bx,\bx')~\left(\altg^*_i(\bx_\alpha,\bx') ~{\mathcal P}_{ij}\right) \right]\right\},\\
&=&  -\frac{1}{T}\int_\odot d\bx~\altg^\dagger_k(\bx,\bx_\beta) ~\left\{\delta{\mathcal L}_{kp}~\left[ \int_\odot d\bx'~{G}_{pj}(\bx,\bx')~\left(\altg^{\dagger}_i(\bx',\bx_\alpha) ~{\mathcal P}_{ij}\right)^* \right]\right\},
\end{eqnarray}
because $\altg_k(\bx_\beta,\bx) \equiv \altg^\dagger_k(\bx,\bx_\beta)$ (from Eqs.~[\ref{def.altg}] and~[\ref{def.altg.2}]) and ${\mathcal P}_{ij}(\omega)$ is real valued.
Recalling equation~(\ref{var_tt}), and transforming to the temporal Fourier domain, we obtain
\begin{equation}
\delta\tau = \frac{1}{2\pi}\int~d\omega~W_{\alpha\beta}^*(\omega)~ \delta\cc_{\alpha\beta}(\omega).\label{freqspace_dt}
\end{equation}
Now, substituting equation~(\ref{freqspace_dt}) into the expression for the misfit (Eq.~[\ref{misfit.orig}] and Eq.~[\ref{temp.misfit}]), we obtain
\begin{equation}
\delta\misfit_1 =  -\sum_q\frac{1}{2\pi T}\int_\odot d\bx\int d\omega~W_{\alpha\beta}^*(\omega)~b^{(n)}_q~\altg^\dagger_k(\bx,\bx_\beta) ~\left\{\delta{\mathcal L}_{kp}~\left[ \int_\odot d\bx'~{G}_{pj}(\bx,\bx')~\left(\altg_i(\bx_\alpha,\bx') ~{\mathcal P}_{ij}\right)^* \right]\right\}_k,\label{misfit.1}
\end{equation}
where some bijective mapping function connects $q$ to the cross-correlation points $(\alpha,\beta)$. We define the adjoint field to be
\begin{equation}
\adj_{\alpha\beta}(\bx) = \bfaltg^\dagger(\bx,\bx_\beta) ~W_{\alpha\beta}^*(\omega)~b^{(n)}_q,\label{adj.eq}
\end{equation}
where subscript $k$ has been dropped from the right side. 
It is important to note that observations have been assimilated into the adjoint field at this stage; thus, kernels that emerge will be functions of measurements. 
A subtlety in implementation arises due to the fact that the filter that takes the raw Green's function to the observable is actually applied on the second
spatial index, $\bx_\beta$. Here we explicitly specify this term
\begin{equation}
\altg_k^\dagger(\bx,\bx_\beta) =[{\mathcal F} * (l_i G^\dagger_{ki})]|_{(\bx,\bx_\beta)} = \int_\odot d\bx{'}~ G^\dagger_{ki}(\bx,\bx{'})~\left[l_i~ {\mathcal F}(\bx_\beta - \bx{'},\omega)\right],\label{altg.dagger}
\end{equation}
where we have assumed a laterally-invariant filter. We arrive at the following adjoint wavefield
\begin{equation}
\adj_{\alpha\beta}(\bx) = \int_\odot d\bx{'}~{\bf G}^\dagger(\bx,\bx{'}) \cdot \boldsymbol{\mathcal M}(\bx{'},\omega).
\end{equation}
The time-domain representation of this field is
\begin{equation}
\adj_{\alpha\beta}(\bx,t) = \int_\odot d\bx{'}\int dt'~{\bf G}^\dagger(\bx,\bx{'},t-t')\cdot \boldsymbol{\mathcal M}(\bx{'}, t'),
\end{equation}
where $\boldsymbol{\mathcal M}$ is a vector whose components are given by
\begin{equation}
{\mathcal M}_i(\bx, \omega) = W_{\alpha\beta}^*(\omega)~b^{(n)}_q\left[l_i~ {\mathcal F}(\bx_\beta - \bx,\omega)\right].\label{adj.source}
\end{equation}

The forward field represents correlations of the wavefield between every point in the domain and the observed pixel $\alpha$, and is calculated in a two-step approach (because of
the presence of two Green's functions). First we compute
the filtered wavefield response to the temporal spectrum of excitation applied at point $\alpha$
\begin{equation}
\bzeta(\bx,\omega) = \int_\odot d\bx'~{\bf G^{\dagger}}(\bx,\bx') \cdot\boldsymbol{\mathcal D},\label{intermed.eq}
\end{equation}
where using equations~(\ref{altg.dagger}) and~(\ref{adj.source}), we define the source 
\begin{equation}
{\mathcal D}_j(\bx,\bx',\omega) = {\mathcal F}(\bx_\alpha - \bx',\omega)\, l_i\, {\mathcal P}_{ij}(\bx,\omega)\label{intermed.source}
\end{equation}
In time domain this equation is
\begin{equation}
\bzeta(\bx,t) = \int_\odot d\bx \int_0^t dt'~{\bf G}^{\dagger}(\bx,\bx',t-t') \cdot\boldsymbol{\mathcal D}(\bx,\bx',t').\label{intermed.eq.time}
\end{equation}
This response in reverse time is applied as a source again, leading to the forward wavefield
\begin{equation}
\fwd_\alpha(\bx) = \int_\odot d\bx'~{\bf G}(\bx,\bx')\cdot \bzeta^*(\bx',\omega),\label{fwd.eq}
\end{equation}
whose time-domain representation is given by
\begin{equation}
\fwd_\alpha(\bx,t) = \int_\odot d\bx'\int_0^t dt'~{\bf G}(\bx,\bx',t-t')\cdot \bzeta(\bx',-t').\label{fwd.eq.time}
\end{equation}

We arrive
at the following interaction integral
\begin{equation}
\delta\misfit_1 = - \sum_{\alpha,\beta} \frac{1}{2\pi T}\int_\odot d\bx\int d\omega~\adj_{\alpha\beta}\cdot(\delta\boldsymbol{\mathcal L}~\fwd_\alpha). \label{misfit_int}
\end{equation}

The second contribution to the variation in misfit may be written as
\begin{eqnarray}
\delta\misfit_2 =  -\sum_q\frac{1}{2\pi T}\int_\odot d\bx\int d\omega~W_{\alpha\beta}^*(\omega)~b^{(n)}_q~\altg^{\dagger *}_k(\bx,\bx_\alpha) ~\left\{\delta\boldsymbol{\mathcal L}^*~\left[ \int_\odot d\bx'~{G}^*_{pj}(\bx,\bx')~\left(\altg_i(\bx_\beta,\bx') ~{\mathcal P}_{ij}(\bx',\omega)\right) \right]\right\}_k,
\end{eqnarray} 
and since all of these functions have purely real temporal representations, integration over frequency allows us to use the relation $\delta\misfit_2^* =  \delta\misfit_2$, whereby
\begin{eqnarray}
\delta\misfit_2 =  -\sum_q\frac{1}{2\pi T}\int_\odot d\bx\int d\omega~W_{\alpha\beta}(\omega)~b^{(n)}_q~\altg^\dagger_k(\bx,\bx_\alpha) ~\left\{\delta\boldsymbol{\mathcal L}~\left[ \int_\odot d\bx'~{G}_{pj}(\bx,\bx')~\left(\altg^\dagger_i(\bx',\bx_\beta) ~{\mathcal P}_{ij}(\bx',\omega)\right)^* \right]\right\}_k,\label{misf.2.eq}
\end{eqnarray} 
which resembles equation~(\ref{misfit.1}), except for the adjoint source now being slightly different and with adjoint and source points exchanged. The algorithm for computing
this second term remains unchanged from that required for the first contribution.
The total misfit variation is given by
\begin{equation}
\delta\misfit = - \sum_{\alpha,\beta} \frac{1}{2\pi T}\int_\odot d\bx\int d\omega~\adj_{\alpha\beta}\cdot(\delta\boldsymbol{\mathcal L}~\fwd_\alpha) + \adj_{\beta\alpha}\cdot(\delta\boldsymbol{\mathcal L}~\fwd_\beta), \label{misfit_tot}
\end{equation}
where wavefields and corresponding sources are read off from the two misfit contributions, $\delta\misfit_1$ (Eq.~[\ref{misfit.1}]) and $\delta\misfit_2$ (Eq.~[\ref{misf.2.eq}]).
In summary, we have deconstructed the meaning of the quantity ``travel time", and expressed it in terms of primitive wavefield descriptors such as Green's functions and sources. Next,
we studied its variation with respect to small perturbations to the wave operator - the first step in determining the Fr\'{e}chet derivative. Having quantified its variation, we decomposed
the Fr\'{e}chet derivative into two constituent wavefields, whose convolution, mediated by an operator, reduces to the sensitivity kernel for that parameter.

\section{Computing Sensitivity Kernels}\label{sens.kernels}
With suitable notation and mathematics in place, we now describe convolution relations between forward and adjoint wavefields which
give sensitivity kernels for various model parameters, such as background flows, sound speed, density and magnetic fields. The latter two, in addition to the equilibrium
equation, determine the corresponding variation in pressure.
We begin with flow kernels; 
changes in isolation to the flow operator are written as $\delta\boldsymbol{\mathcal L} = - 2 i \omega\rho \vel\cdot\bnabla$.
Substituting this into equation~(\ref{misfit_int}), we obtain
\begin{eqnarray}
\delta\misfit_1 &=& 2i \sum_{\alpha,\beta}\frac{1}{2\pi T}\int_\odot d\bx \int d\omega~\omega\rho~\adj_{\alpha\beta}\cdot(\vel\cdot\bnabla)\fwd_\alpha\nonumber\\
&=& \int_\odot d\bx~ \vel \cdot {\bf K}^{(1)}_{\vel},
\end{eqnarray}
where,
\begin{equation}
{\bf K}^{(1)}_{\vel}(\bx) = 2i\rho \sum_{\alpha,\beta}\frac{1}{2\pi T} \int d\omega~\omega~(\bnabla\fwd_\alpha)\cdot\adj_{\alpha\beta}.
\end{equation}
Alternately, written in time domain, the flow sensitivity kernel becomes
\begin{equation}
{\bf K}^{(1)}_{\vel}(\bx) = -2\sum_{\alpha,\beta} \frac{1}{T}\int dt~\rho[\bnabla\partial_t\fwd_\alpha(t)]\cdot\adj_{\alpha\beta}(-t),\label{timed}
\end{equation}
where for sake of convenience, we do not explicitly state the $\bx$ dependence of the two fields.
If we were to compute forward and adjoint fields based on equations~(\ref{fwd.eq}) and~(\ref{adj.eq}), then interaction~(\ref{timed}) between forward and time-reversed adjoint fields 
gives us the desired sensitivity kernel. The second contribution to misfit (and therefore the kernel) must be computed and added to equation~(\ref{timed}), i.e.,
\begin{equation}
{\bf K}_{\vel} = {\bf K}^{(1)}_{\vel} + {\bf K}^{(2)}_{\vel},
\end{equation}
where
\begin{equation}
{\bf K}^{(2)}_{\vel}(\bx) = -2 \sum_{\alpha,\beta}\frac{1}{T} \int dt~\rho~(\bnabla\partial_t\fwd_\beta(t))\cdot\adj_{\beta\alpha}(-t).
\end{equation}

Next, we consider perturbations to sound speed, $\delta\boldsymbol{\mathcal L} = - \bnabla(\rho\delta c^2~\bnabla\cdot )$. Substituting this in equation~(\ref{misfit_int}), we have
\begin{eqnarray}
\delta\misfit_1 &=& \sum_{\alpha,\beta}\frac{1}{2\pi T}\int_\odot d\bx\int d\omega~ \adj_{\alpha\beta}\cdot\bnabla(\rho\delta c^2~\bnabla\cdot\fwd_\alpha ) \\
&=& \sum_{\alpha,\beta}\frac{1}{2\pi T}\int_\odot d\bx\int d\omega~ \bnabla\cdot(\rho\delta c^2~\adj_{\alpha\beta}\bnabla\cdot\fwd_\alpha ) - \rho\delta c^2~\bnabla\cdot\adj_{\alpha\beta}~\bnabla\cdot\fwd_\alpha.
\end{eqnarray}
The first term reduces to a surface integral at the domain boundaries and may therefore be dropped (having assumed homogeneous boundary conditions as $\bx \rightarrow \infty$). The sound-speed
kernel reduces to
\begin{eqnarray}
\delta\misfit_1 &=& \int_\odot d\bx~{\delta\ln c^2}~ K^{(1)}_{c^2},\\
K^{(1)}_{c^2}(\bx) &=& -\rho c^2\sum_{\alpha,\beta}\frac{1}{2\pi T} \int d\omega~\bnabla\cdot\adj_{\alpha\beta}~\bnabla\cdot\fwd_\alpha.
\end{eqnarray}
Alternately, in time domain, the sound-speed kernel is obtained upon computing
\begin{equation}
K^{(1)}_{c^2}(\bx) = -\rho c^2\sum_{\alpha,\beta}\frac{1}{T} \int dt~\bnabla\cdot\adj_{\alpha\beta}(-t)~\bnabla\cdot\fwd_\alpha(t).\label{kernel.ss}
\end{equation}

In order to derive kernel expressions for magnetic field, density and pressure, which are additionally constrained by the 
equilibrium equation, we consider small perturbations around~(\ref{equili}), namely
\begin{equation}
\bnabla \delta p = \delta\rho {\bf g} + \rho\, \delta{\bf g} + (\bnabla\curl\delta{\bf B})\curl{\bf B} + (\bnabla\curl{\bf B})\curl\delta{\bf B}.\label{perturb.equili}
\end{equation}
We may ignore perturbations to the gravitational field arising from surface phenomena, such as sunspots or flows in the convection zone, because an overwhelming fraction of solar mass
is concentrated within the radiative interior. We have
\begin{eqnarray}
\delta\misfit_1 &=& \frac{1}{2\pi T}\int_\odot d\bx~\adj_{\alpha\beta}\cdot\bnabla(\fwd_\alpha\cdot\bnabla\delta p),\\
 &=& - \frac{1}{2\pi T}\int_\odot d\bx~\bnabla\cdot\adj_{\alpha\beta} \fwd_\alpha\cdot\bnabla\delta p,\\
  &=& - \frac{1}{2\pi T}\int_\odot d\bx~\bnabla\cdot\adj_{\alpha\beta} \fwd_\alpha\cdot[\delta\rho\, {\bf g} + (\bnabla\curl\delta{\bf B})\curl{\bf B} + (\bnabla\curl{\bf B})\curl\delta{\bf B}].\label{misf.pre}
\end{eqnarray}

Terms involving density and magnetic field in the misfit expression for pressure (Eq.~[\ref{misf.pre}]) are absorbed into kernel expressions for the former two quantities respectively. The density kernel follows
\begin{eqnarray}
\delta\misfit_1 &=& \int_\odot d\bx~K^{'(1)}_{\rho}~\delta\rho\\
K^{'(1)}_{\rho} &=& -\sum_{\alpha,\beta}\frac{1}{2\pi T}\int d\omega~\left[-\omega^2~\adj_{\alpha\beta}\cdot\fwd_\alpha - i\omega\Gamma~\adj_{\alpha\beta}\cdot\fwd_\alpha  +~  c^2\bnabla\cdot\adj_{\alpha\beta}\bnabla\cdot\fwd_\alpha\right.\nonumber\\
 && \left.  -\, \fwd_\alpha\cdot\bnabla {\bf g}\cdot\adj_{\alpha\beta} -{\bf g}\cdot(\fwd_\alpha\cdot\bnabla\adj_{\alpha\beta} + \fwd_\alpha\bnabla\cdot\adj_{\alpha\beta})\right],
\end{eqnarray}
which in time domain is
\begin{eqnarray}
K^{'(1)}_{\rho} &=& -\sum_{\alpha,\beta}\frac{1}{T}\int dt~\left\{\adj_{\alpha\beta}(-t)\cdot\partial^2_t\fwd_\alpha(t) + \adj_{\alpha\beta}(-t)\cdot[\Gamma *\partial_t\fwd_\alpha(t)] +  ~c^2\bnabla\cdot\adj_{\alpha\beta}(-t)\bnabla\cdot\fwd_\alpha(t)  \right.\nonumber\\
 && \left. -\, \fwd_\alpha(t)\cdot\bnabla {\bf g}\cdot\adj_{\alpha\beta}(-t) -{\bf g}\cdot[\fwd_\alpha(t)\cdot\bnabla\adj_{\alpha\beta}(-t) + \fwd_\alpha(t)\bnabla\cdot\adj_{\alpha\beta}(-t)]\right\}
\end{eqnarray}
This is the same as the kernel expression in, e.g., equation~(55) of \citet{liu_tromp_08} with their rotation terms and gravitational potential variations (their $\psi$, $\delta\phi$, and $\delta g_0$ respectively) set to zero.
\citet{zhu09}, who term this the {\it impedance kernel} ($K'_\rho$), showed that it is primarily sensitive to reflection zones and discontinuities. It therefore remains to be seen whether much information about density variations in the solar interior may be extracted from the wavefield.

Dropping the assumption of translation invariance allows us to derive simple vector expressions for magnetic field kernels (see appendix~\ref{magkernels}).
The effect of small deviations from a background
field on travel times is given by the following expression
\begin{eqnarray}
\delta\misfit_1 &=& \int_\odot d\bx~{\bf K}^{(1)}_{\bB} \cdot \delta\bB,\\
{\bf K}^{(1)}_{\bB} &=& \sum_{\alpha,\beta}\frac{1}{2\pi T}\int d\omega~\bnabla\curl[\bnabla\curl(\fwd_\alpha\curl\bB) \curl\adj_{\alpha\beta}] + \left\{\bnabla\curl[\adj_{\alpha\beta} \curl (\bnabla\curl\bB)]\right\}\curl \fwd_\alpha \nonumber\\
&& +  \adj_{\alpha\beta}\curl\{\bnabla\curl[\bnabla\curl(\fwd_\alpha \curl \bB)]\} + \fwd_\alpha\curl\{\bnabla\curl[\bnabla\curl(\adj_{\alpha\beta}\curl\bB)] \}\nonumber\\
&& + \bnabla\curl[{\bf B} \curl (\fwd_\alpha\bnabla\cdot\adj_{\alpha\beta} )] + \bnabla\cdot\adj_{\alpha\beta} \fwd_\alpha\curl[\bnabla\curl{\bf B}],
\end{eqnarray}
which in time domain is
\begin{eqnarray}
{\bf K}^{(1)}_{\bB} &=& \sum_{\alpha,\beta}\frac{1}{T}\int dt~\bnabla\curl[\bnabla\curl(\fwd_\alpha(t)\curl\bB) \curl\adj_{\alpha\beta}(-t)] + \left\{\bnabla\curl[\adj_{\alpha\beta}(-t) \curl (\bnabla\curl\bB)]\right\}\curl \fwd_\alpha(t) \nonumber\\
&& +  \adj_{\alpha\beta}(-t)\curl\{\bnabla\curl[\bnabla\curl(\fwd_\alpha(t) \curl \bB)]\} + \fwd_\alpha(t)\curl\{\bnabla\curl[\bnabla\curl(\adj_{\alpha\beta}(-t)\curl\bB)] \}\nonumber\\
&& + \bnabla\curl[{\bf B} \curl (\fwd_\alpha(t)\bnabla\cdot\adj_{\alpha\beta}(-t) )] + \bnabla\cdot\adj_{\alpha\beta}(-t) \fwd_\alpha(t)\curl[\bnabla\curl{\bf B}].
\end{eqnarray}
For inversions constrained by equilibrium equation~(\ref{equili}), we have arrived at kernels for four independent model parameters, namely: sound speed, velocity, density, and magnetic field. 

What is the connection between kernels derived here and those computed by e.g., \citet{birch}? Translation invariance implies that everywhere in the domain of interest, differences between
predicted and measured travel times are small and that perturbations to the background state are weak. 
Dividing out the $\Delta\tau_q$ term (setting it to some constant value) from expressions for the kernel and misfit,
we arrive at the classical linear helioseismic forward problem
\begin{equation}
\delta\tau = \int_\odot d\bx ~{\bf K}(\bx)\cdot\delta {\bf q}(\bx),
\end{equation}
where $\delta{\bf q}$ is a perturbation of interest.

\section{Performing Inversions, Computational Algorithm, \& Cost}
In this section, we describe how the adjoint technique may be applied efficiently to perform large-scale inversions using helioseismic data.
We begin with the concept of an {\it event kernel}, discussed in e.g., \citet{bamberger82, igel96, tromp05,CarlTape09}, the focus of the inverse procedure. Consider a seismic event (i.e., a source at) $\alpha$ whose signature
is recorded at some $N$ locations. In a computational sense, the predicted wavefield generated by this source event is encoded in the forward wavefield, while
observations are assimilated into the adjoint wavefield. The expense involved in computing event kernels
scales linearly with number of sources $\alpha$, independently of the number of observation locations.
This is because (as will be shown here) all $N$ observations may simultaneously be injected at corresponding station locations
to produce the adjoint field. In the translationally-invariant helioseismology case, the event kernel may be obtained by summing up $N$ appropriately rotated and translated kernels, each weighted by the relevant travel-time shift
\begin{equation}
K_{\alpha}(\bx) = \sum_{\beta=1}^N \Delta\tau_{\alpha\beta} K_{\alpha\beta}(\bx).
\end{equation}
We formulate algorithmic details associated with incorporating large numbers of observations into the inversion \citep[see also][]{tromp10}.
Let us choose $M$ master pixels, which are correlated with signals measured at $N$ pixels, i.e., $MN+M(M-1)/2$ correlations
in total, a number that scales as $O(MN)$ since $M\ll N$. The associated misfit may be written as
\begin{equation}
\delta\misfit = -\sum_{\beta=1}^N\sum_{\alpha=1}^M\frac{1}{2\pi T} \int_\odot d\bx\int  d\omega \left(\adj_{\alpha\beta}\cdot\delta\boldsymbol{\mathcal L}\fwd_\alpha + \adj_{\beta\alpha}\cdot\delta\boldsymbol{\mathcal L}\fwd_\beta\right).
\end{equation}
We attempt to moderate computational cost by absorbing the summation over $N$ into forward and adjoint sources, suitably redefined. The first contribution to
misfit may be rewritten as
\begin{equation}
\delta\misfit_1 = -\sum_{\beta=1}^N\sum_{\alpha=1}^M \frac{1}{2\pi T}\int_\odot d\bx\int d\omega~\adj_{\alpha\beta}\cdot\delta\boldsymbol{\mathcal L}\fwd_\alpha =-\sum_{\alpha=1}^M \frac{1}{2\pi T}\int_\odot d\bx\int d\omega ~{\baradj}_{\alpha}\cdot\delta\boldsymbol{\mathcal L}\fwd_\alpha,\label{contrib.1}
\end{equation}
where the adjoint source and wavefield are given by
\begin{eqnarray}
{\mathcal M}_i(\bx) &=& l_i\sum_{\beta=1}^N {\mathcal F}(\bx_\beta-\bx,\omega)~W^*_{\alpha\beta}~b^n_q,\label{adjsource.2}\\
{\baradj}_{\alpha}(\bx) &=& \int_\odot d\bx'~{\bf G}(\bx,\bx')\cdot\boldsymbol{\mathcal M}(\bx',\omega),\label{adj.eq.2}
\end{eqnarray}
and the forward wavefield is as stated in equation~(\ref{fwd.eq}). All $N$ cross correlations of slave pixels with master-pixel $\alpha$ are subsumed into one adjoint calculation.
Computationally, this is accomplished by constructing adjoint source~(\ref{adjsource.2}) as a sum over all slave pixels that are correlated with $\alpha$.
The partial event kernel $K^{(1)}_\alpha$ may be computed using the wavefields in equation~(\ref{contrib.1}) together with kernel expressions stated
in the preceding section.
The second contribution requires some manipulation and redefinitions, namely
\begin{eqnarray}
\delta\misfit_2 = -\sum_{\beta=1}^N\sum_{\alpha=1}^M \frac{1}{2\pi T}\int_\odot d\bx\int d\omega~\adj_{\beta\alpha}\cdot\delta\boldsymbol{\mathcal L}\fwd_\beta &=& -\sum_{\alpha=1}^M \frac{1}{2\pi T}\int_\odot d\bx\int d\omega~\baradj_{\alpha}\cdot\delta\boldsymbol{\mathcal L}\barfwd_\alpha,
\label{misf.2.event.kernel}\\
{\mathcal M}_i(\bx) &=& l_i {\mathcal F}(\bx_\alpha-\bx,\omega)\label{adjsource.3},\\
{\baradj}_{\alpha}(\bx) &=& \int_\odot d\bx'~{\bf G}(\bx,\bx')\cdot \boldsymbol{\mathcal M}(\bx',\omega),\label{adj.eq.3}\\
\boldsymbol{\mathcal{D}}_\alpha(\bx,\bx',\omega) &=& \sum_{\beta=1}^N W_{\alpha\beta}\, b^{(n)}_q~{\mathcal F}(\bx'-\bx_\beta)\, {\bf\hat l}\cdot\boldsymbol{\mathcal P}(\bx,\omega),\label{inter.inv}\\
\bar\bzeta_\alpha(\bx,\omega) &=& \int_\odot d\bx{'}~{\bf G^\dagger}(\bx,\bx{'})\cdot~ \boldsymbol{\mathcal D}_\alpha(\bx,\bx{'},\omega),\label{fwd.inv}\\
\barfwd_\alpha &=&\int_\odot d\bx'~{\bf G}(\bx,\bx')\cdot\bar\bzeta_\alpha(\bx',\omega).\label{adj.inv}
\end{eqnarray}
The second contribution is constructed by interacting the two wavefields according to equation~(\ref{misf.2.event.kernel}) and added to $K^{(1)}_\alpha$ to complete the calculation of the full event kernel.
Thus the vast number of observations of the solar wavefield may all be assimilated into the inversion but with a finite $O(M)$ number of calculations. The 
$M$ master pixels may be chosen to ensure greatest coverage within the region of interest, whose locations could be decided by criteria such as
maximizing the sum of distances between point pairs. The algorithm, depending on whether sensitivity kernels are being computed or inversions
are performed may be stated in the following manner:

\begin{itemize}
\item {\it Master Pixels}: Choose a set of $M$ master ($\alpha$) and $N$ slave ($\beta$) pixels with $M\ll N$. For instance, a constellation of points surrounding a sunspot or active region.
\item {\it Intermediate Wavefield} ($\bzeta, \bar\bzeta$): If the intent is to compute kernels, source~(\ref{intermed.source}) is applied (Eq.~[\ref{intermed.eq}]) and the resulting wavefield is 
saved at all points where the wave excitation source is non-zero. Alternately, when performing inversions, two types of sources, given by~(\ref{intermed.source}) and~(\ref{inter.inv}), must
be applied. 
Because this wavefield is used to drive the forward simulation, it must be saved at a sufficient number of temporal points. This does not demand large 
storage requirements since only 2D slices are written out (for all practical purposes, wave excitation occurs at one depth). The driving source for the calculation of a sensitivity kernel
is given by~(\ref{intermed.eq}) and for the event kernel~(\ref{inter.inv}). This is termed the {\it generating wavefield} by \citet{tromp10}.
\item {\it Forward wavefield} ($\fwd, \barfwd$): Driven by the time-reversed intermediate wavefield displacement ($\bzeta,\bar\bzeta$) injected at the nominal excitation depth, with 
the specific choice of sources dependent on whether an event kernel~(\ref{fwd.inv}) or a sensitivity kernel~(\ref{fwd.eq}) is being computed.
The 3D wavefield is saved at a cadence of 30 seconds (Nyquist frequency of 16.66 mHz).
This is termed the {\it ensemble forward wavefield} by \citet{tromp10}.
\item {\it Adjoint Source} ($\boldsymbol{\mathcal M}$): The time-history of the forward wavefield extracted at the observation height  is filtered according to equation~(\ref{def.altg})
and time series at all slave pixels are isolated. These form the predicted limit cross correlations for those point pairs. We now determine the adjoint
source according to equations~(\ref{adj.source}),~(\ref{adjsource.2}), or~(\ref{adjsource.3}) as the case may be (i.e., computing kernels between a point pair or performing an inversion
using large numbers of observations). Note this is the stage where observations are assimilated into the inversion.
\item {\it Adjoint Wavefield \& Partial Kernels} ($\adj, \baradj$): The former is evaluated according to equations~(\ref{adj.eq}),~(\ref{adj.eq.2}), or~(\ref{adj.eq.3}) as the case may be. 
The 3D adjoint wavefield is 
saved at the same cadence as that of the forward. We may then compute kernels 
according to interaction integral~(\ref{misfit_int}). 
Each sensitivity or event kernel has two contributions which must be added together.
\item{\it Temporal length \& Computational domain size}: Simulations must be run for at least as long as it takes for waves to arrive from the farthest contributing source to the observation points. 
The farther the source is, the greater the effects of damping and geometric spreading and thus the contribution of a source diminishes with distance from observation points.
\item{\it Storage Cadence}: Five to ten points per temporal wavelength is a reasonable rule of thumb. This is done in order to maximize the accuracy in evaluating the interaction integral~(\ref{misfit_int}) while not placing
unnecessary demands on storage. Of course, this step may be obviated if one were to apply the algorithm of \citet{liu_tromp_08}.
\item{\it Boundary conditions}: Highly-absorbent boundary conditions are recommended in order that waves that have propagated out do not return to the region of interest.
\end{itemize}

For a fixed resolution and temporal extent of the calculation, both storage and computational expense increase linearly with the number of master pixels, i.e., computations scale as
 $O(5M)$ Green's function calculations, where $M$ is the number of master pixels. This is because the intermediate wavefield needs only be computed for a time extent $T/2$ whereas the forward and adjoint wavefields
must be calculated over a temporal length $T$. 
Storage cost scales approximately as $O(4M)$. The power of this technique is twofold, firstly
in being able to compute all kernels relevant to a given measurement simultaneously from the adjoint and forward wavefields, and secondly, in assimilating as many
observations as desired in order to perform the inversion. 

Rapid convergence, i.e., reduction in misfit, is a desirable quality of an inverse technique. Two well-known drawbacks of the steepest descent method are that it converges very slowly
for problems where the condition number is large and the convergence rate is very sensitive to the local step-size ($\epsilon$ in Eq.~[\ref{update.eqs}]). 
A much more popular and powerful technique is the conjugate-gradient method, which utilizes misfit gradients at current and previous iterations in 
order to determine the directionality and magnitude of the step to be taken. Preconditioning gradients in order to reduce the condition number
and improve convergence characteristics is also a typically-employed procedure. We shall not describe these issues in any greater detail at present
but merely note their importance and that they need be addressed in any inverse procedure. For an in-depth discussion of these topics, see, e.g., \citet{tape07}.

An important aspect of the outcome of an inversion relates to uniqueness. Because this is an optimization problem, the solution may be trapped in a local minimum.
One may attempt to avoid this pitfall by adopting the so-called multi-scale approach \citep[e.g.,][]{bunks95, sirgue04, ravaut04, fichtner09} which involves taking the following precautionary steps:
\begin{itemize}
\item Choosing a ``good" initial model is crucial since meaningless local optima may attract and trap the solution. In the case of sunspots, one may construct 3D models that are
constrained by the surface field.
\item Employ travel times of long-wavelength waves (i.e., high phase speeds) that are primarily sensitive to coarse-grained features of the object in question and iteratively refine the model by 
gradually incorporating travel times of smaller wavelength waves (lower phase speeds). 
\item Use different types of measurements, i.e., a variety of time-distance averaging geometries, frequency and phase-speed filters, in the misfit function.
\item Ensure a good match between simulations and observations at the photospheric level (e.g., photospheric sunspot magnetic fields or Doppler measurements of surface supergranulation).
\end{itemize}

\section{Flow and sound-speed kernels}
We use the Seismic Propagation through Active Regions and Convection (SPARC) code, developed by \citet{hanasoge_thesis, Hanasoge_couvidat_2008}. A highly-efficient absorption method 
termed the convolutional perfectly-matched layer formulated for stratified 
environments \citep[][]{hanasoge_2010} is applied at all boundaries. A domain of size $250 \times 250 \times 35~ {\rm Mm}^3$ is chosen, where the first two
dimensions are horizontal and the third depth. The box straddles the photosphere, extending from 34 Mm below to 1 Mm above. The grid consists of $384\times384\times300$ points,
ensuring a horizontal resolution of $660$ km. Vertical grid spacing decreases smoothly from about 250 km at the bottom of the box to around 27 km
at the photosphere and above, so designed as to maintain constant acoustic travel time between adjacent pairs of points. 

\begin{figure}[!ht]
\centering
\epsscale{1.0}
\plotone{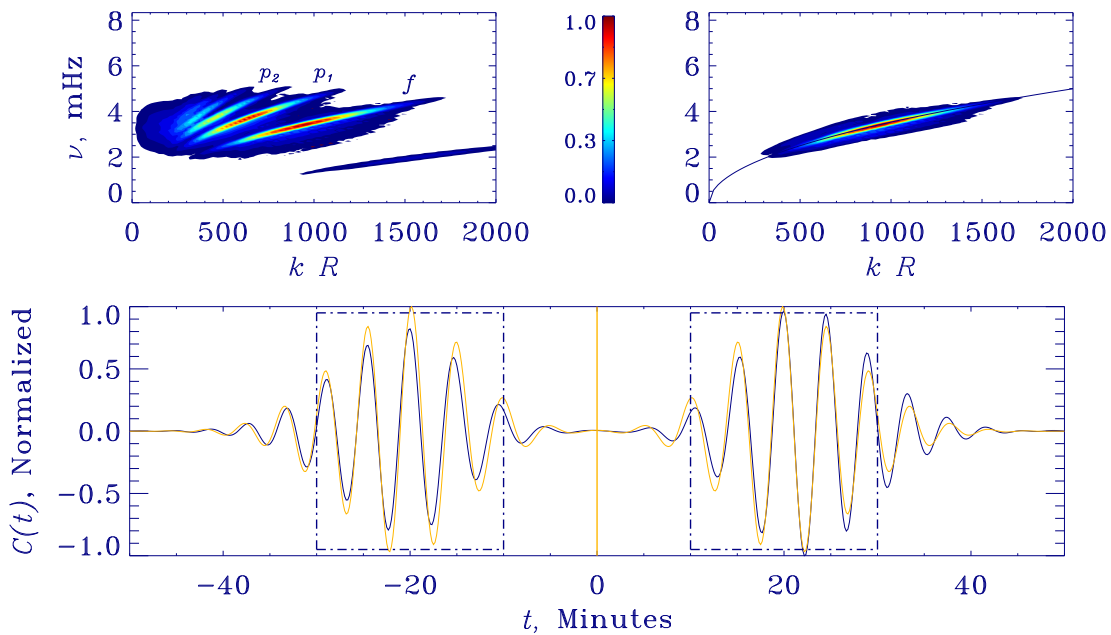}
\caption{Normalized pre- and post-filtered power spectra of the $z$ component of $\eta$ on the upper panels and the $f$-mode dispersion relation $\omega = \sqrt{gk}$ overplotted (dark line; right).
Below is shown the normalized predicted $f$-mode limit cross correlation $\cc_{\alpha\beta}(t)$ of the wavefield measured at a pair of points with separation distance $|\alpha-\beta| = 10~ {\rm Mm}$. Overplotted
with the thin line is the ``exact" cross correlation, estimated by inverse Fourier transforming the power spectrum (Eq.~[\ref{exact.cc}]).
The amplitudes of the negative and positive branches are slightly different but their phases match well. The dot-dash boxes around branches of the limit cross correlation denote the 
chosen temporal window ($f(t)$ in Eq.~[\ref{def.weight}]).}\label{pspecfig}
\end{figure}

In Figure~\ref{pspecfig}, power spectra of pre- and post-filtered intermediate wavefields (vertical component of $\bzeta$) 
are shown; we isolate the $f$-mode for this calculation. The predicted limit cross correlation, $\cc_{\alpha\beta}(t)$,
obtained by filtering the forward wavefield and extracting the time series at the receiver is also shown. The positive and negative branches differ slightly in amplitude but show
good phase agreement.

\begin{figure}[!ht]
\centering
\epsscale{1.0}
\plotone{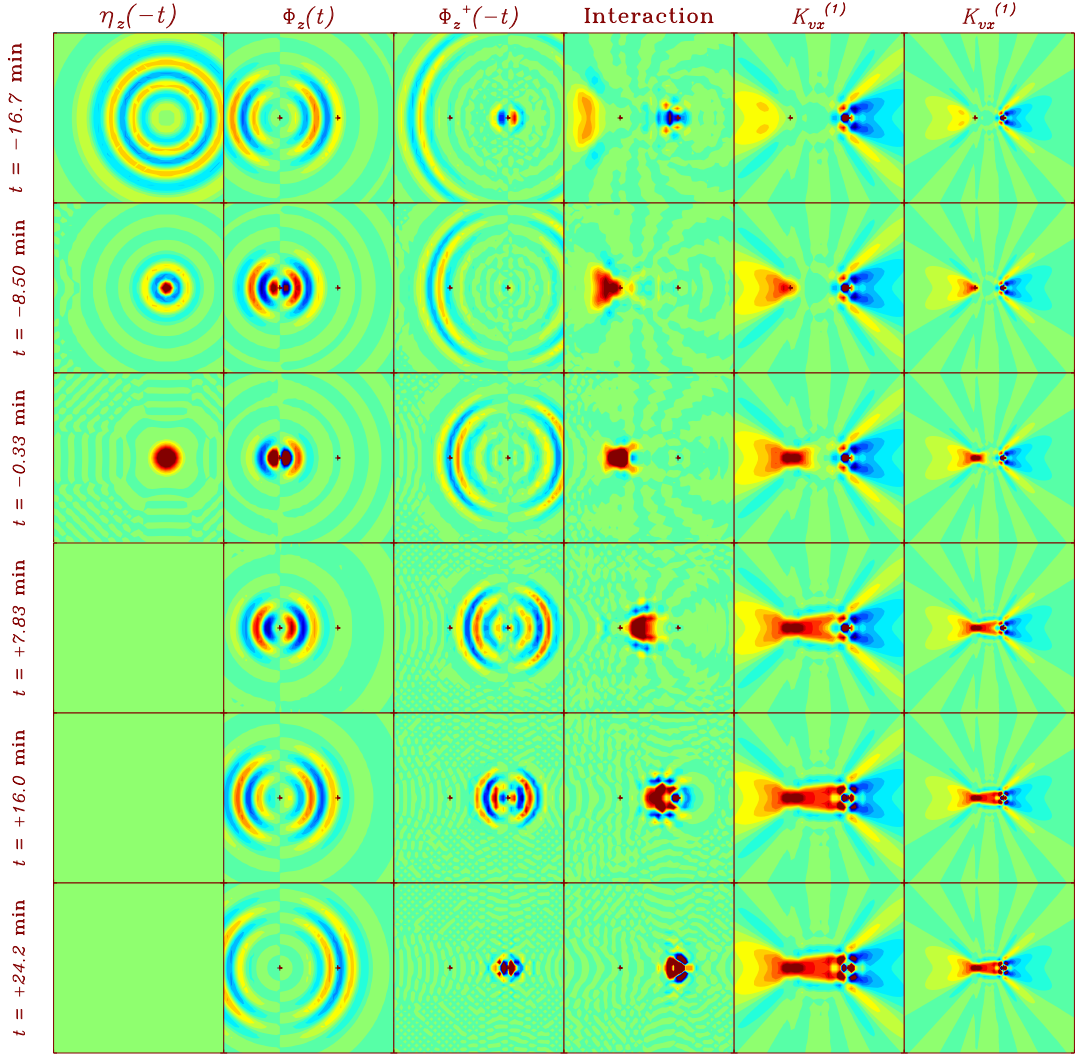}
\caption{Snapshots in time of various wavefields, interaction integral, and kernel zoomed in and out \citep[akin to Figures 2 and 3 of ][]{tromp10}. 
The first three columns show vertical components of $\bzeta, \fwd_\alpha, \adj_{\alpha\beta}$, and the next two display
the interaction integral and horizontal flow kernel, $\misfit_1, K^{(1)}_{v_{x}}$ with the final column showing a zoomed out picture of the kernel. 
The adjoint field $\adj_{\alpha\beta}$ and spectral response wavefield $\eta$ are shown in reverse time in order to highlight the computational algorithm: (1) we time reverse
$\eta$ and feed into the forward calculation and (2) the kernel is calculated via a convolution between the forward and adjoint.
The two marks denote locations of source (left) and receiver. Total solar time of the spectral response calculation 
($\bzeta$; left column) is 2.5 hrs, and forward and adjoint are 5 hrs each. Note that the forward field (second column) is centered around the source point while the adjoint (third column) is centered around
the receiver.
}\label{snapshots}
\end{figure}

We display actual wavefields and demonstrate the process of computing kernels graphically in Figure~\ref{snapshots}. 
The first column shows snapshots of the intermediate wavefield $\bzeta$ forced by a 
source at $\alpha$ at three time instants ---
this wavefield is filtered, time reversed and fed into the code as a source for the forward wavefield, $\fwd_\alpha$, seen in the second column. The adjoint source is computed using the predicted
limit cross correlation that is derived from the forward wavefield and used to drive the adjoint wavefield $\adj_{\alpha\beta}$ (where $\beta$ is the receiver),
depicted in reverse time in the third column at a number of instants. The final two
columns show the interaction integral and stages in the construction of the partial kernel. This entire process must be repeated with $\beta$ as source and $\alpha$ the receiver and
its contribution must be added to the partial kernel obtained previously (shown in Figure~\ref{kernels_slices}).

Cuts through kernels are shown in Figure~\ref{kernels_slices}. The upper three panels show partial contributions and full kernels, at a depth $z = -0.5$ Mm; vertical
cuts through the $y=0$ center-line for $K_{v_{x}}$ and $K_{v_{z}}$ are displayed on the fourth panel. 
Only the $f$-mode contributes to the kernel as evidenced by the constancy in sign of the kernel as a function of depth. The $x-$ and $y-$ (anti-) symmetries
of kernels are as expected \citep[see e.g.,][]{birch_gizon_07}. 
The two faint horizontal lines seen at $z = -0.2$ Mm and $z=0.2$ Mm in the vertical cuts of the $x-$ and $z-$ kernels
correspond to the excitation depth
and observation height respectively; the intermediate wavefield is computed with a source at the former depth and the cross-correlation and adjoint sources are injected
at the latter height. The integral of the $x$-flow kernel may be directly estimated from the power spectrum --- the two values agree to within a few percent (see appendix~{\ref{kern.integ}).
\begin{figure}[!ht]
\centering
\epsscale{1.0}
\plotone{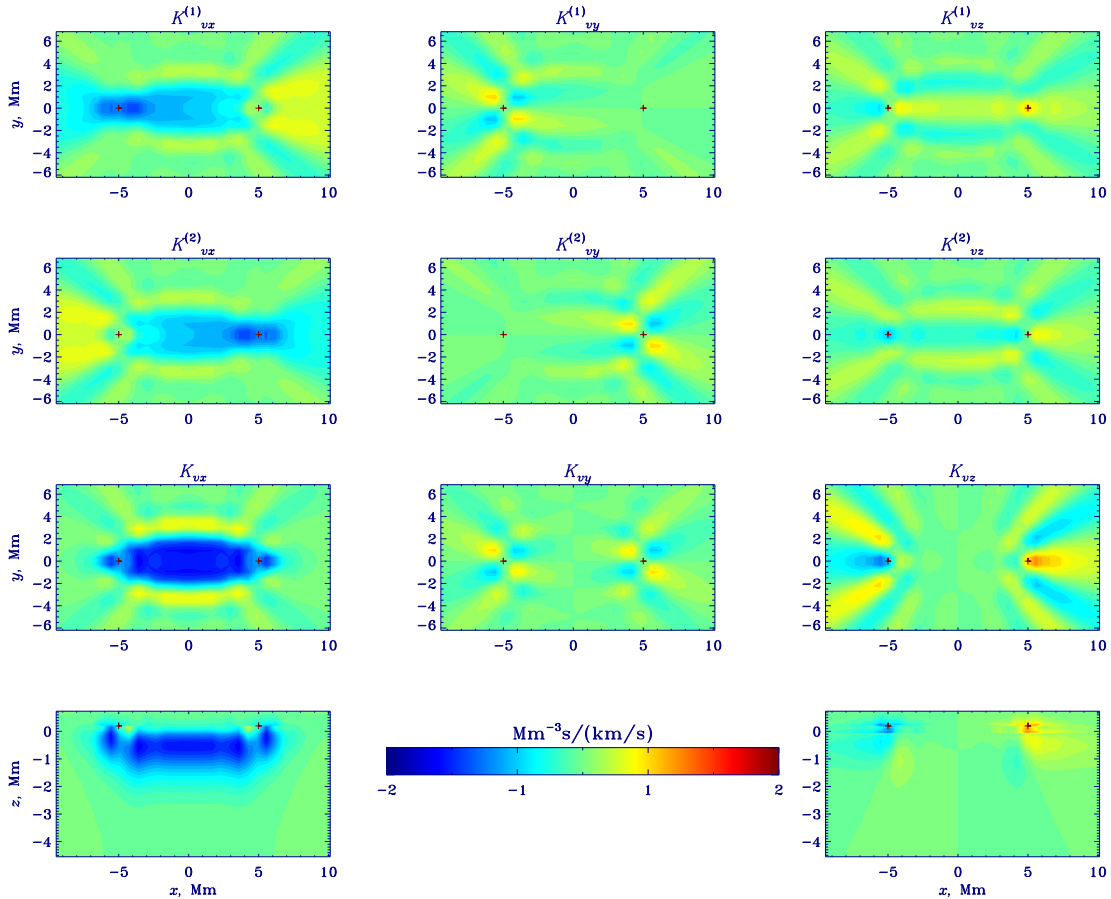}
\caption{Partial contributions to flow kernels (first two rows) and their sum (third row), displayed at a depth of $z=-0.54$ Mm. 
A vertical cut through the $K_{v_{x}}$ and $K_{v_{z}}$ kernels 
along the $y=0$ center line is shown on the fourth row. These are computed around the translationally-invariant polytropic background 
described in appendix~\ref{full.stratification}. Symmetries and magnitudes of the kernels are in line with expectation \citep{birch_gizon_07}. Note that there is an extra factor of time
in the dimension of the flow kernels that arises from assimilating observed travel times into the kernels.}\label{kernels_slices}
\end{figure}

We also compute the sound-speed kernel for the mean travel-time measured using the $p_1$ ridge. Mean travel times are measured according to equation~(\ref{weight.ss}). The intermediate, forward,
and adjoint simulations are performed and the interaction of the latter two is computed in accordance with equation~(\ref{kernel.ss}). The filtered power spectrum and limit cross correlation
for the measurement are shown in Figure~\ref{pspec.p1}. The positive and negative branches are slightly phase shifted, suggesting the requirement of a larger computational domain. The kernel
for this measurement is shown in Figure~\ref{kernelc}. The raypath corresponding to 10 Mm angular (horizontal) distance is also plotted for reference; note similarities to sound-speed kernels computed by \citet{birch}.
\begin{figure}[!ht]
\centering
\epsscale{1.0}
\plotone{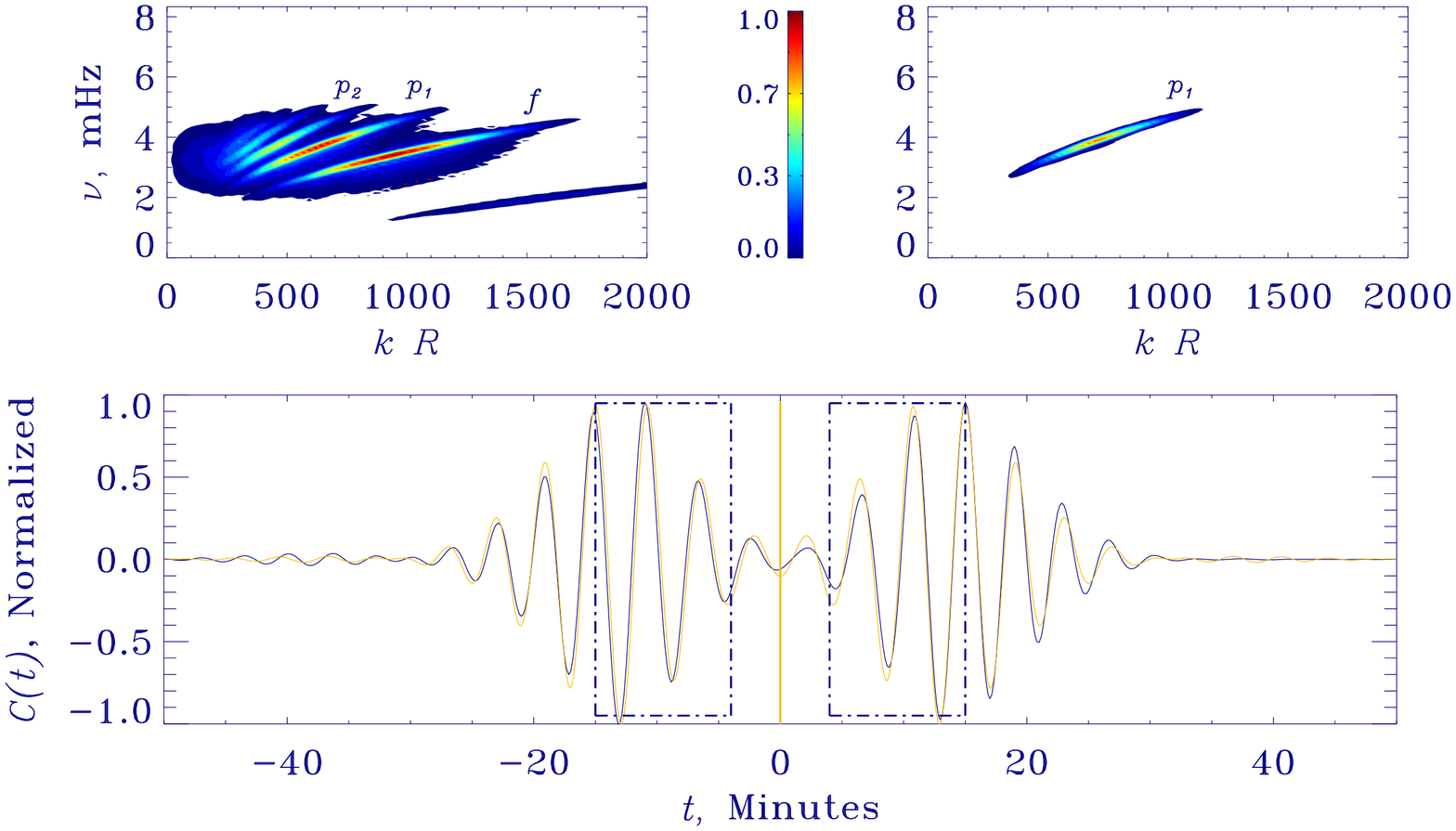}
\caption{Filtered $p_1$ spectrum and limit cross correlation for a pair of antennae 10 Mm apart. The thin line depicts the expected cross correlation, estimated from Fourier transforming the power spectrum (Eq.~[\ref{exact.cc}]).
}\label{pspec.p1}
\end{figure}

\begin{figure}[!ht]
\centering
\epsscale{1.0}
\plotone{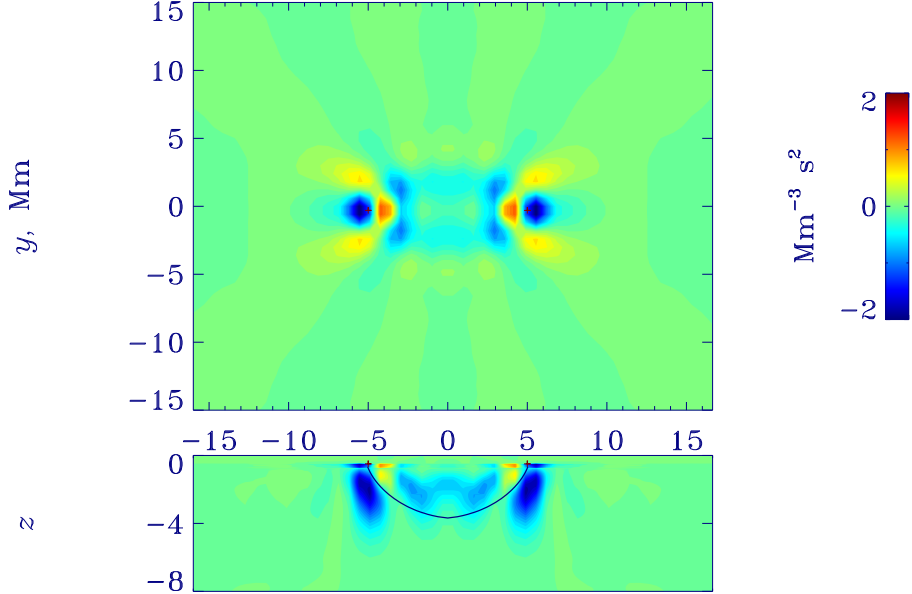}\vspace{-2cm}
\caption{Sound-speed kernel for a mean travel-time measurement using a $p_1$ mode.  The antennae are separated by 10 Mm. The solid line
denotes the ray path. A double-bounce $p$ wave, whose ray travel time is approximately 15 minutes (and therefore likely in the temporal window), is also noticeable.
}\label{kernelc}
\end{figure}

\section{Conclusions}
We present a general algorithm to compute sensitivity kernels and discuss its application in
inversions for solar structure and dynamics.
One may envisage using this method to compute kernels for the spherical Sun, for problems that require non-linear 
inverse techniques such as sunspots and supergranules, and perhaps move towards the greater goal
of full-waveform inversion where properties derived from cross correlations, such as amplitude, may also be accounted
for. Computational improvements, such as the ability to recover identical positive and negative branches of the cross correlation
for a laterally-invariant background are required in order that amplitude information may be effectively used. Further, one must also consider 
the fundamental problem of convective instability in near-surface layers of models of solar stratification before this technique
of computing kernels becomes directly applicable to the Sun.

\acknowledgements
The computing required to bring this technique to fruition was mostly performed on the 
Pleiades cluster at NASA Ames. S.M.H. acknowledges support from NASA grant NNX11AB63G and the DLR grant ``German Data Center for SDO". 
This work contributes to the deliverables identified in FP7 European Research
Council grant agreement 210949, ``Seismic Imaging of the Solar Interior", to
PI L. Gizon (Milestone 2 and contribution towards Milestones 4 and 5).
Many thanks
to Yang Luo (Princeton) for suggestions that improved readability of the manuscript.

\appendix
\section{Seismic Reciprocity}\label{sec.reciprocity}
We recall the helioseismic wave operator defined in equation~(\ref{waveop.helio}).
The operator may be split into two parts, Hermitian ${\mathcal H}$ and anti-Hermitian ${\mathcal H}^\dagger$, where the former satisfies the following relation
\begin{equation}
\int_\odot d\bx~\bxi_A\cdot{\mathcal H}\bxi_B = \int_\odot d\bx~\bxi_B\cdot{\mathcal H}\bxi_A.\label{relation1}
\end{equation}
The proof of the self-adjointness of the ideal MHD equations with damping and no background flow is fairly intricate and will not be repeated here. For a thorough demonstration, please refer
to e.g., \citet{goedbloed2004}.
The only anti-Hermitian part of equation~(\ref{waveop.helio}) contains the background velocity term
\begin{equation}
-2i\omega\int_\odot d\bx~\bxi_A\cdot~({\rho\vel\cdot\bnabla})\bxi_B = 2i\omega\int_\odot d\bx~\bxi_B\cdot~({\rho\vel\cdot\bnabla})\bxi_A,\label{relation2}
\end{equation}
where the sign of the two integrals upon switching states $A$ and $B$ is reversed. Green's theorem (also Eq.~[\ref{greenseq}]) tells us
\begin{equation}
[({\mathcal H} -2i\omega\vel\cdot\bnabla){\bf G}(\bx,\bx_A)]_{ip} = \delta_{ip}~\delta(\bx - \bx_A)\label{gov.eq.a}.
\end{equation}
In order to demonstrate reciprocity, we consider another wave state due to a source $B$, whose Green's function is given by
\begin{equation}
[({\mathcal H} + 2i\omega\vel\cdot\bnabla){\bf G}^\dagger(\bx,\bx_B)]_{iq} = \delta_{iq}~\delta(\bx - \bx_B)\label{gov.eq.b}.
\end{equation}
Now consider forming the following representation ${G}^\dagger_{qi}(\bx,\bx_B)\times$(\ref{gov.eq.a}) $ - ~{G}_{pi}(\bx,\bx_A)\times$(\ref{gov.eq.b}) and integrating over all space. We have
\begin{eqnarray}
&&\int_\odot d\bx~{G}^\dagger_{qi}(\bx,\bx_B)~[({\mathcal H} -2i\omega\vel\cdot\bnabla){\bf G}(\bx,\bx_A)]_{ip} - 
{G}_{pi}(\bx,\bx_A)~[({\mathcal H}+2i\omega\vel\cdot\bnabla){\bf G}^\dagger(\bx,\bx_B)]_{iq} \nonumber\\
&&= \int_\odot d\bx~{G}^\dagger_{qi}(\bx,\bx_B)~\delta_{ip}~\delta(\bx - \bx_A) - {G}_{pi}(\bx,\bx_A)~\delta_{iq}~\delta(\bx - \bx_B).\label{representation1}
\end{eqnarray}
Equations~(\ref{relation1}) and~(\ref{relation2}) imply the left-hand side of~(\ref{representation1}) is zero. Thus we arrive at the seismic reciprocity relation for helioseismic waves
\begin{equation}
{G}^\dagger_{qp}(\bx_A,\bx_B,\omega)={G}_{pq}(\bx_B,\bx_A,\omega),
\end{equation}
where $G^\dagger$ is Green's function for an identical wave operator, except with flows reversed in sign. The adjoint operator $\boldsymbol{\mathcal L}^\dagger$ is therefore
\begin{eqnarray}
\boldsymbol{\mathcal L}^\dagger\bxi&=&-\omega^2\rho\bxi -i\omega\rho\Gamma\bxi +2i\omega\rho\vel\cdot\bnabla\bxi - \bnabla(c^2\rho\bnabla\cdot\bxi  +  \bxi\cdot\bnabla p) - \bnabla\cdot(\rho\bxi){\bf g} \nonumber\\
&-&\left[\bnabla\curl\bB \curl \bnabla\curl(\bxi \curl \bB) + \{\bnabla\curl[\bnabla\curl(\bxi \curl \bB)]\} \curl \bB\right].
\end{eqnarray}

\section{Magnetic field kernels}\label{magkernels}
The perturbed magnetic operator is described by
\begin{eqnarray}
\delta\boldsymbol{\mathcal L}\,\bxi &=& -(\bnabla\curl\delta\bB) \curl [\bnabla\curl(\bxi \curl \bB)] - \{\bnabla\curl[\bnabla\curl(\bxi \curl \delta\bB)]\} \curl \bB\nonumber\\
&-& (\bnabla\curl\bB) \curl [\bnabla\curl(\bxi \curl \delta\bB)] - \{\bnabla\curl[\bnabla\curl(\bxi \curl \bB)]\} \curl \delta\bB.
\end{eqnarray}
The variation in the misfit is given by
\begin{eqnarray}
\delta\misfit_1 &=& \frac{1}{T}\sum_{\alpha,\beta}\int_\odot d\bx\int d\omega~ \adj_{\alpha\beta}\cdot\bigg\{ (\bnabla\curl\delta\bB) \curl [\bnabla\curl(\fwd_\alpha \curl \bB)] +\{\bnabla\curl[\bnabla\curl(\fwd_\alpha \curl \delta\bB)]\} \curl \bB\nonumber\\
&+& (\bnabla\curl\bB) \curl [\bnabla\curl(\fwd_\alpha \curl \delta\bB)] + \{\bnabla\curl[\bnabla\curl(\fwd_\alpha \curl \bB)]\} \curl \delta\bB\bigg\}.
\end{eqnarray}
In order to free the $\delta\bB$ from the confines of the differential curl operator, we make use of the following vector identities,
\begin{eqnarray}
{\bf a}\cdot({\bf b} \curl {\bf c}) &=& {\bf c}\cdot({\bf a} \curl {\bf b}) = {\bf b}\cdot({\bf c} \curl {\bf a})\\
{\bf a}\cdot\bnabla\curl{\bf b} &=& {\bf b}\cdot\bnabla\curl{\bf a} - \bnabla\cdot({\bf a} \curl{\bf b} ),
\end{eqnarray}
and the fact that
\begin{equation}
\int_\odot d\bx~\bnabla\cdot({\bf a} \curl{\bf b} ) = 0,
\end{equation}
due to the homogeneous upper boundary conditions we employ.
Taking the first term, we have
\begin{eqnarray}
\adj_{\alpha\beta}\cdot\bigg[ (\bnabla\curl\delta\bB) \curl \bnabla\curl(\fwd_\alpha \curl \bB) \bigg] &=& (\bnabla\curl\delta\bB)\cdot\bigg\{[\bnabla\curl(\fwd_\alpha \curl \bB)]\curl \adj_{\alpha\beta}\bigg\},\\
\int_\odot d\bx~(\bnabla\curl\delta\bB)\cdot\bigg[\bnabla\curl(\fwd_\alpha \curl \bB)\curl \adj_{\alpha\beta}\bigg ] &=& \int_\odot d\bx~\delta\bB\cdot\bigg\{\bnabla\curl[\bnabla\curl(\fwd_\alpha \curl \bB)\curl \adj_{\alpha\beta}]\bigg \},
\end{eqnarray}
and now the second,
\begin{eqnarray}
\adj_{\alpha\beta}\cdot\bigg[ \{\bnabla\curl[\bnabla\curl(\fwd_\alpha \curl \delta\bB)]\} \curl \bB \bigg] &=& \{\bnabla\curl[\bnabla\curl(\fwd_\alpha \curl \delta\bB)]\}\cdot(\bB\curl \adj_{\alpha\beta}),\\
\int_\odot d\bx~\{\bnabla\curl[\bnabla\curl(\fwd_\alpha \curl \delta\bB)]\}\cdot(\bB\curl \adj_{\alpha\beta}) &=& \int_\odot d\bx~ [\bnabla\curl(\fwd_\alpha \curl \delta\bB)] \cdot \bnabla\curl(\bB\curl \adj_{\alpha\beta}),\\
\int_\odot d\bx~ [\bnabla\curl(\fwd_\alpha \curl \delta\bB)] \cdot \bnabla\curl(\bB\curl \adj_{\alpha\beta}) &=& \int_\odot d\bx~(\fwd_\alpha \curl \delta\bB) \cdot  \bnabla\curl[\bnabla\curl(\bB\curl \adj_{\alpha\beta})],\\
(\fwd_\alpha \curl \delta\bB) \cdot  \bnabla\curl[\bnabla\curl(\bB\curl \adj_{\alpha\beta})] &=& \delta\bB\cdot\bigg\{\bnabla\curl[\bnabla\curl(\bB\curl \adj_{\alpha\beta})]\curl\fwd_\alpha \bigg\},
\end{eqnarray}
followed by the third
\begin{eqnarray}
\adj_{\alpha\beta}\cdot\bigg [(\bnabla\curl\bB) \curl \bnabla\curl(\fwd_\alpha \curl \delta\bB)\bigg ] &=& \bnabla\curl(\fwd_\alpha \curl \delta\bB)\cdot\bigg [\adj_{\alpha\beta}\curl(\bnabla\curl\bB)\bigg ],\\
\int_\odot d\bx~\bnabla\curl(\fwd_\alpha \curl \delta\bB)\cdot\bigg [\adj_{\alpha\beta}\curl(\bnabla\curl\bB)\bigg ] &=& \int_\odot d\bx~(\fwd_\alpha \curl \delta\bB)\cdot\bigg\{\bnabla\curl[\adj_{\alpha\beta}\curl(\bnabla\curl\bB)]\bigg\},\\
(\fwd_\alpha \curl \delta\bB)\cdot\bigg\{\bnabla\curl[\adj_{\alpha\beta}\curl(\bnabla\curl\bB)]\bigg\} &=& \delta\bB\cdot\bigg\{ \bnabla\curl[\adj_{\alpha\beta}\curl(\bnabla\curl\bB)] \curl\fwd_\alpha\bigg\},
\end{eqnarray}
and finally, the simplest of them all
\begin{eqnarray}
\adj_{\alpha\beta}\cdot\bigg\{ \bnabla\curl[\bnabla\curl(\fwd_\alpha \curl \bB)] \curl \delta\bB \bigg\}  &=& \delta\bB\cdot\bigg\{\adj_{\alpha\beta}\curl \bnabla\curl[\bnabla\curl(\fwd_\alpha \curl \bB)]  \bigg\}.
\end{eqnarray}

There are other terms that arise from perturbing the equilibrium equation~(\ref{perturb.equili}). These are
\begin{eqnarray}
-\frac{1}{T}\int_\odot d\bx~(\bnabla\cdot\adj_{\alpha\beta})\, \fwd_\alpha\cdot[(\bnabla\curl\delta{\bf B})\curl{\bf B} + (\bnabla\curl{\bf B})\curl\delta{\bf B}].
\end{eqnarray}

Expanding on the first,
\begin{eqnarray}
(\bnabla\cdot\adj_{\alpha\beta}) \,\fwd_\alpha\cdot(\bnabla\curl\delta{\bf B})\curl{\bf B} &=&  (\bnabla\curl\delta{\bf B})\cdot[{\bf B} \curl (\fwd_\alpha\bnabla\cdot\adj_{\alpha\beta} )],\\
\int_\odot d\bx~(\bnabla\curl\delta{\bf B})\cdot[{\bf B} \curl (\fwd_\alpha\bnabla\cdot\adj_{\alpha\beta} )] &=& \int_\odot d\bx~\delta{\bf B}\cdot\{\bnabla\curl[{\bf B} \curl (\fwd_\alpha\bnabla\cdot\adj_{\alpha\beta} )]\},
\end{eqnarray}
and the second may be manipulated so
\begin{eqnarray}
(\bnabla\cdot\adj_{\alpha\beta}) \,\fwd_\alpha\cdot(\bnabla\curl{\bf B})\curl\delta{\bf B} = \delta{\bf B}\cdot[\bnabla\cdot\adj_{\alpha\beta} \fwd_\alpha\curl(\bnabla\curl{\bf B})].
\end{eqnarray}

\section{Background Stratification}\label{full.stratification}
For the sub-surface layers, we use the following polytropic stratification prescription:
\begin{eqnarray}
p(z) &=& p_{\rm poly} \left(1 - \frac{z}{z_f}\right)^{m+1},\nonumber\\
\rho(z) &=& \rho_{\rm poly} \left(1 - \frac{z}{z_f}\right)^{m},\nonumber\\
g &=& \frac{m+1}{z_f} \frac{p_{\rm poly}}{\rho_{\rm poly}}\,,\label{polytrope.eq}\\
c(z) &=& \sqrt{\frac{\frac{m+1}{m} p_{\rm poly}}{\rho_{\rm poly}}\left(1 - \frac{z}{z_f}\right)}\,\,.\nonumber\\
\end{eqnarray}
We set $p_{\rm poly} = 1.178 \times 10^5~{\rm dynes~ cm^{-2}}, \rho_{\rm poly} = 3.093 \times 10^{-7}~{\rm g~ cm^{-3}}, z_f = - 0.450~{\rm Mm},$ and $m=2.150$. Note also that $z$ is the height, i.e., the atmospheric
layers are described by $z>0$ and vice versa; $z=0$ is the fiducial surface.

Waves propagating toward the surface in the Sun are reflected by a fluctuation in the background density gradient. This reflection zone is located at a 
height of $z_r\sim-0.050$ Mm below the surface. We mimic this by attaching the above polytropic stratification with an overlying isothermal layer. This patching 
results in a fluctuation in the density gradient,
leading to an acoustic cut-off frequency of approximately 5.4 mHz, similar to the solar value. For $z < z_M$, we apply the above prescription. For $z \ge z_r$,
we use the following equations
\begin{eqnarray}
p(z) &=& p_{\rm iso} \exp\left[\frac{z_r - z}{H}\right],\nonumber\\
p_{\rm iso} &=& p_{\rm poly} \left(1 - \frac{z_r}{z_f}\right)^{m+1},\nonumber\\
\rho(z) &=& \rho_{\rm iso} \exp\left[\frac{z_r - z}{H}\right],\nonumber\\
\rho_{\rm iso} &=& \rho_{\rm poly} \left(1 - \frac{z_r}{z_f}\right)^{m},\\
H &=& \frac{p_{\rm iso}}{g \rho_{\rm iso}}\,.\nonumber
\end{eqnarray}
The relations for $\rho_{\rm iso}$ and $p_{\rm iso}$ arise from the requirement of continuity of the pressure and density at the matching point between the polytropic and isothermal
layers. The relation for $H$ is a consequence of enforcing hydrostatic balance. This model is truncated at $z = -34$ Mm (lower boundary; polytrope) and $z=+1$ Mm (upper boundary; 
isothermal layer).

\section{Units of Wavefields, Conventions, and Definitions}\label{conventions}
We apply the following Fourier transform convention
\begin{eqnarray}
\int_{-\infty}^\infty dt~e^{i\omega t}~ g(t) &=& g(\omega) ,\\
\int_{-\infty}^\infty dt~e^{i\omega t} &=& 2\pi~\delta(\omega),\\
\frac{1}{2\pi}\int_{-\infty}^{\infty} d\omega~e^{-i\omega t}~ g(\omega) &=& g(t),\\
\int_{-\infty}^\infty d\omega~e^{-i\omega t} &=&  2\pi~\delta(t),
\end{eqnarray}
where the Fourier-transform pair $g(t), g(\omega)$ are written similarly for convenience. 
The equivalence between cross-correlations and convolutions in the Fourier and temporal domain are written so
\begin{equation}
h(t) = \int_{-\infty}^{\infty} dt'~ f(t')~ g(t+t') \Longleftrightarrow h(\omega) = f^*(\omega)~g(\omega), \\
\end{equation}
\begin{equation}
h(t) = \int_{-\infty}^{\infty} dt'~ f(t')~ g(t-t') \Longleftrightarrow h(\omega) =  f(\omega)~g(\omega).
\end{equation}
The following relationship also holds (for real functions $f(t), g(t)$)
\begin{equation}
\int_{-\infty}^\infty dt~f(t)~g(t) = \frac{1}{2\pi}\int_{-\infty}^{\infty} d\omega~f^*(\omega)~g(\omega) = \frac{1}{2\pi}\int_{-\infty}^{\infty} d\omega~f(\omega)~g^*(\omega).
\end{equation}
We now describe the physical units of various quantities (indicated by square brackets around a given variable)
\begin{itemize}
\item $[\delta(\bx)]\equiv {\rm Mm^{-3}}~~~~~(\int_\odot d\bx~\delta(\bx) = 1)$
\item $[\delta(t)]\equiv {\rm s^{-1}}~~~~~(\int dt~\delta(t) = 1)$
\item $[\boldsymbol{\mathcal L}]\equiv {\rm g\cdot Mm^{-3}\cdot ~s^{-2}}~~~~~(\boldsymbol{\mathcal L} \sim \rho\omega^2)$ 
\item $[{\bf G}]\equiv {\rm s \cdot g^{-1}}~~~~~[\boldsymbol{\mathcal L}{\bf G} = \delta(\bx-\bx')\delta(t-t')]$ 
\item $[{\bf S}]\equiv {\rm g\cdot Mm^{-3}\cdot~Mm\cdot s^{-2}}~~~~~~[\boldsymbol{\mathcal L}\bxi(\bx,t) = {\bf S}(\bx,t)]$
\item $[\boldsymbol{\altg}]\equiv {\rm g^{-1}}~~~~~[\int_\odot d\bx'~dt'~\boldsymbol\altg(\bx,\bx',t-t')\cdot~{\bf S}(\bx',t') = \phi(\bx,t)]$ 
\item $[{\mathcal F}]\equiv{\rm Mm^{-3}~s^{-2}}~~~~~~[\altg_j(\bx,\bx{''},t) = \int~dt'~d\bx'~ {\mathcal F}(\bx',t')~l_i~{ G}_{ij}(\bx-\bx',\bx{''},t-t')]$
\item $[\boldsymbol{\mathcal P}]\equiv {\rm g^2\cdot Mm^{-3}\cdot~Mm^2\cdot s^{-3}}~~~~~~[{\mathcal P}_{ij}(\bx,\omega) \delta(\bx-\bx') = \langle S_i(\bx,\omega) S_j^*(\bx',\omega) \rangle]$
\item $[\bzeta]\equiv {\rm g\cdot~Mm^{-1}\cdot s^{-2}}~~~~~~[=\int dt'~\boldsymbol{\altg}(\bx,\bx_\alpha,t-t')\cdot\boldsymbol{\mathcal P}(\bx,t')]$
\item $[\mathcal M_i]\equiv {\rm Mm^{-5}\cdot s^2}~~~~~~[=\int dt'~l_ib_q~W_{\alpha\beta}(t'+t)~{\mathcal F}(\bx_\beta-\bx',t')]$
\item $[\fwd]\equiv {\rm Mm^2}~~~~~~[=\int dt'~d\bx'~{\bf G}(\bx,\bx',t-t')\cdot{\bzeta}(\bx',t')]$
\item $[\adj]\equiv {\rm g^{-1}\cdot Mm^{-2}\cdot s^{4}}~~~~~~[=\int dt'~d\bx'~{\bf G}(\bx,\bx',t-t')\cdot\boldsymbol{\mathcal M}(\bx',t')]$
\item $[{\bf K}_{\vel}] \equiv {\rm Mm^{-4}\cdot s^{3}}~~~~~~[ = \frac{1}{T}\int dt~\rho [\bnabla\partial_t\fwd(t)]\cdot\adj(-t)]$
\end{itemize}

We use equation~(4) from \citet{gizon04} in order to define the weight function $W_{\alpha\beta}(t)$ for the differential flow measurement
\begin{equation}
W_{\alpha\beta}(t) = -{\dot\cc}_{\alpha\beta}(t)\,\frac{ f(t) + f(-t)}{\Delta t\sum_{t'}  f(t') \left[{\dot\cc}_{\alpha\beta}(t')\right]^2  }\,,\label{def.weight}
\end{equation}
where $\Delta t$ is the temporal rate at which the cross correlations are sampled, $f(t)$ is a window, 
and the difference travel time $\delta\tau$ is given by
\begin{equation}
\delta\tau = \int dt~ W_{\alpha\beta}(t)~\delta\cc_{\alpha\beta}(t).
\end{equation}
Note that since we compute difference travel times, $W_{\alpha\beta}$ is an odd function of time whose
Fourier transform is therefore purely imaginary. For the sound-speed kernel, we measure mean travel times, defined as
\begin{equation}
W_{\alpha\beta}(t) = -\frac{1}{2}\,{\dot\cc}_{\alpha\beta}(t)\,\frac{ f(t) - f(-t)}{\Delta t\sum_{t'}  f(t') \left[{\dot\cc}_{\alpha\beta}(t')\right]^2  }\,.\label{weight.ss}
\end{equation}

\section{Validation}\label{kern.integ}
We perform validation tests in order to test the quality of computed kernels and limit cross correlations.
\subsection{Classical-tomographic sound-speed kernel}
As a simple test, we compute a single-source sound-speed kernel between a pair of points located 15 Mm apart. The source point is forced
with the function shown in the upper-most panel of Figure~\ref{waveforms}; this calculation forms the forward wavefield. 
The seismogram at the receiver 15 Mm away is shown in the middle panel where the dot-dash lines denote the temporal window applied to
isolate the first arrival. The adjoint source, the time-reversed windowed seismogram, is applied at the receiver. The kernel is subsequently 
calculated according to equation~(\ref{kernel.ss}) and is shown in Figure~\ref{kernels_classical}.

Consider the travel time of a ray propagating along path $\Gamma$
\begin{equation}
\tau = \int_{\mathcal R} \frac{ds}{c},
\end{equation}
where $s$ is length measured along raypath ${\mathcal R}$.
Fermat assures us that the raypath is invariant under small perturbations of sound speed. Therefore the perturbation in travel time due to spatially constant $\delta c/c$ is given by
\begin{equation}
\delta\tau = -\frac{\delta c}{c}\int_\Gamma \frac{ds}{c} = -\tau \frac{\delta c}{c}.
\end{equation}
The sound-speed kernel must therefore satisfy (having divided out $\Delta\tau$),
\begin{equation}
\delta\tau = \int_\odot d\bx~ \frac{\delta c^2}{c^2} K_{c^2}(\bx) \approx -\tau \frac{\delta c}{c},
\end{equation}
or
\begin{equation}
\int_\odot d\bx~K_{c^2}(\bx) \approx -\frac{\tau}{2}.
\end{equation}
We find the integral of the kernel to be $-192.88$ s, which compares well with half the travel time, $-190$ s.

\begin{figure}[!ht]
\centering
\epsscale{1.0}
\plotone{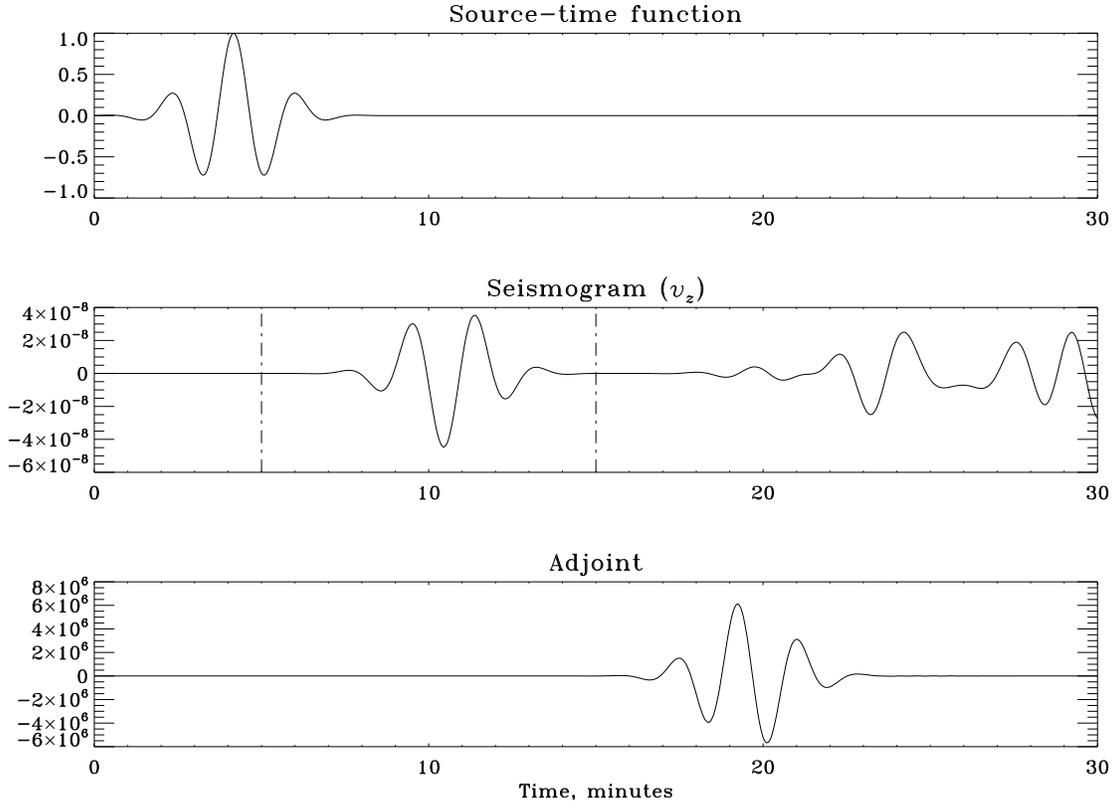}
\caption{Source-time function, receiver seismogram, and adjoint source involved in the computation of a single-source sound-speed kernel \citep[e.g.,][]{tromp05}.
Vertical and horizontal cuts are shown. The dot-dash lines in the seismogram show the temporal window applied to isolate the first arrival.}\label{waveforms}
\end{figure}

\begin{figure}[!ht]
\centering
\epsscale{1.0}
\plotone{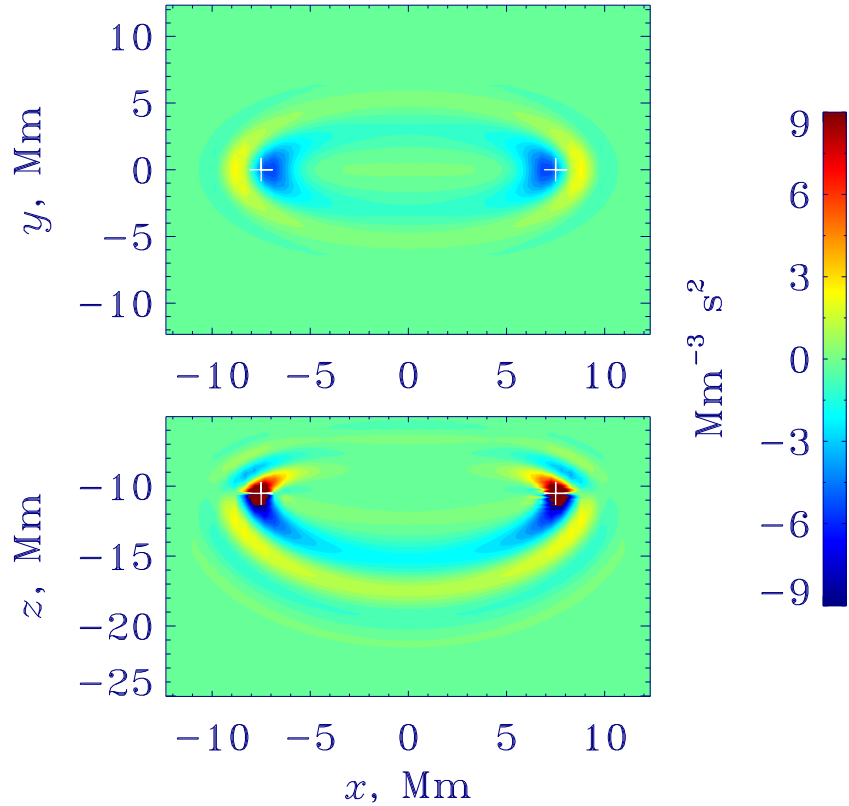}
\caption{A single-source sound-speed kernel \citep[e.g.,][]{gizon02, tromp05}. Vertical and horizontal cuts are shown. The symbols
denote source (left) and receiver positions. The integral
of the kernel is -192.88 s, compared to a half wave travel time of -190 s.}\label{kernels_classical}
\end{figure}

\subsection{Cross correlations}
The filtered cross correlation for a translationally-invariant background model may be written as
\begin{equation}
\cc(\boldsymbol\Delta,\omega) = \int d\bk~|\phi(\bk,\omega){\mathcal F}(\bk,\omega)|^2~e^{i\bk\cdot\boldsymbol\Delta},\label{exact.cc}
\end{equation}
where $\boldsymbol\Delta$ is the vector connecting two observation points. Thus we may estimate the cross correlation between a given
pair of points by inverse-Fourier transforming the power spectrum. In Figures~(\ref{pspecfig}) and~(\ref{pspec.p1}), we compare computed and spectrally-estimated
cross correlations and find some differences that likely arise from the finite-size of the horizontal domain and absorption boundary conditions that 
dissipate high-group-speed (low-frequency) waves which reach boundaries first.

\subsection{Flow-kernel Integral}
We introduce a spatially-uniform 0.1 km/s $x$-directed flow (i.e., everywhere in the domain) and the corresponding travel-time shift using 
equations~(\ref{def.weight}) and~(\ref{var_tt}) is $-\,9.6$ s. The
change in cross correlation for this background model is displayed in Figure~\ref{deltacc_flow}. The flow-kernel integral is
\begin{equation}
\delta\tau = v_{x}\int_\odot d\bx~K_{v_{x}}(\bx) = 0.1 \int_\odot d\bx~K_{v_{x}}(\bx) = - 10.1 \,{\rm s}\,.
\end{equation}
As expected, integrals of $K_{v_{y}}$ and $K_{v_{z}}$ are zero \citep{birch_gizon_07}. 

\begin{figure}[!ht]
\centering
\epsscale{1.0}
\plotone{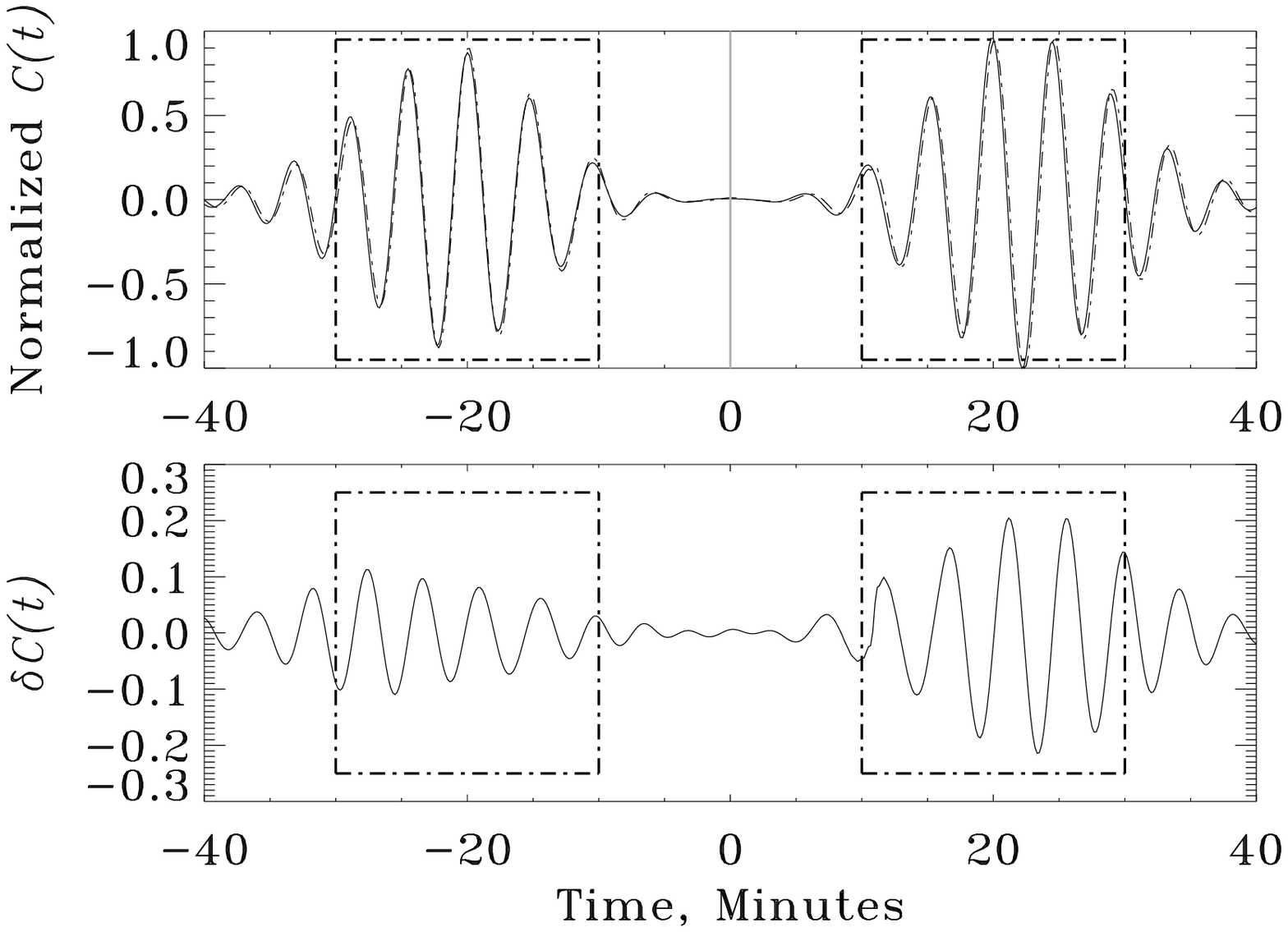}
\caption{The upper panel displays cross correlations between a point pair 10 Mm apart, corresponding to background models with no flow (solid line) and a constant $x-$directed flow of magnitude 0.1 km/s. 
The related travel-time shift, computed using~(\ref{def.weight}) and~(\ref{var_tt}), is $-\,9.6$ s, which implies a kernel integral of $-\, 96$ s.
}\label{deltacc_flow}
\end{figure}

\subsection{Integral of the multiple-source sound-speed kernel}
We introduce a spatially uniform 1\% perturbation to $c^2$ (i.e., everywhere in the domain) and the corresponding travel-time shift computed using equations~(\ref{weight.ss}) and~(\ref{var_tt})
is $-\, 1.98$ s. The 
change in cross correlation for this slightly-altered background model is displayed in Figure~\ref{deltacc_ssp}.

\begin{figure}[!ht]
\centering
\epsscale{1.0}
\plotone{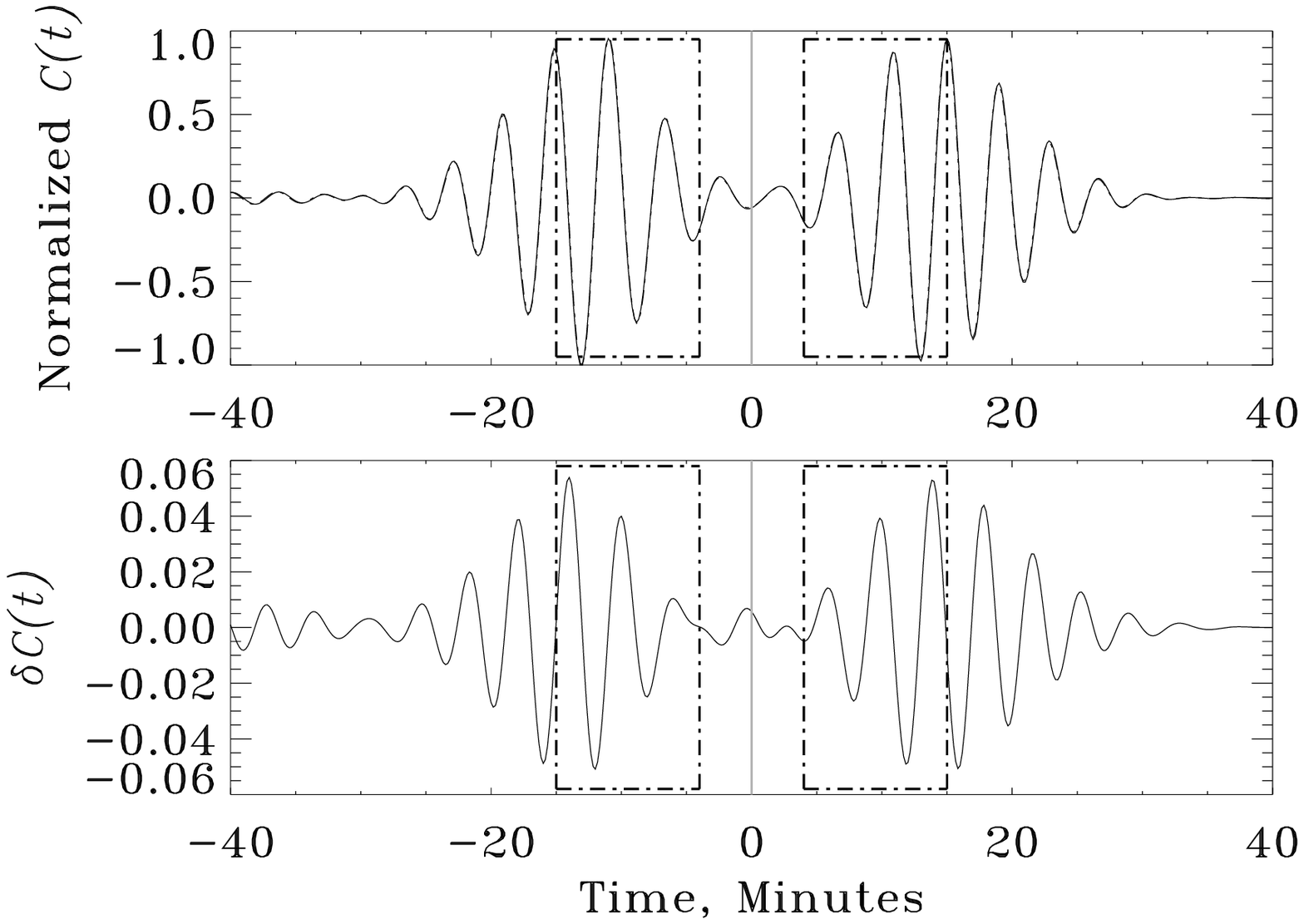}
\caption{The upper panel displays cross correlations between a point pair 10 Mm apart, corresponding to background models with sound-speed distributions of $c^2$ and $1.01 c^2$. The two cross correlations fall almost
on top of other and their difference, only visible on the lower panel, is on the order of a few percent. The related travel-time shift, computed using~(\ref{weight.ss}) and~(\ref{var_tt}), is $-\,1.98$ s, implying a kernel
integral of $-198$ s.
}\label{deltacc_ssp}
\end{figure}

The expected travel-time shift for such a perturbation is given by the following integral
\begin{equation}
\delta\tau = \int_\odot d\bx~K_{c^2}(\bx) \frac{\delta c^2}{c^2} = 0.01 \int_\odot d\bx~K_{c^2}(\bx) = -1.75\, {\rm s}\,,
\end{equation}
where the kernel is displayed in Figure~\ref{kernelc}

\bibliographystyle{apj}
\bibliography{adjoint}

\end{document}